\documentclass[aip,amsmath,amssymb,reprint,preprintnumbers]{revtex4-2}

\usepackage{graphicx}
\usepackage{dcolumn}
\usepackage{bm}

\usepackage[utf8]{inputenc}
\usepackage[T1]{fontenc}
\usepackage{mathptmx}
\usepackage{etoolbox}

\makeatletter
\def\@email#1#2{%
 \endgroup
 \patchcmd{\titleblock@produce}
  {\frontmatter@RRAPformat}
  {\frontmatter@RRAPformat{\produce@RRAP{*#1\href{mailto:#2}{#2}}}\frontmatter@RRAPformat}
  {}{}
}%
\makeatother

\newcommand\trick[1]{}

\usepackage[x11names]{xcolor}
\usepackage{tikz}
\usepackage{comment}
\usetikzlibrary{decorations.pathreplacing,decorations.markings}
\usepgflibrary{arrows}
\usepgflibrary[arrows]
\usetikzlibrary{arrows}
\usetikzlibrary[arrows]
\usetikzlibrary{positioning}

\usepackage{subsupscripts}
\usepackage{braket}
\usepackage{hyperref}
\usepackage{amsthm}
\usepackage[makeroom]{cancel}

\makeatletter

\@addtoreset{equation}{section}
\makeatother
\newcommand{\ba}{\begin{eqnarray}}
\newcommand{\ea}{\end{eqnarray}}

\newcommand{\bpsi}{\bar\psi}

\newcommand{\para}[1]{
\vspace{3mm}

\noindent\textbf{#1}  }

\usetikzlibrary{decorations.pathreplacing,decorations.markings}
\tikzset{
  on each segment/.style={
    decorate,
    decoration={
      show path construction,
      moveto code={},
      lineto code={
        \path [#1]
        (\tikzinputsegmentfirst) -- (\tikzinputsegmentlast);
      },
      curveto code={
        \path [#1] (\tikzinputsegmentfirst)
        .. controls
        (\tikzinputsegmentsupporta) and (\tikzinputsegmentsupportb)
        ..
        (\tikzinputsegmentlast);
      },
      closepath code={
        \path [#1]
        (\tikzinputsegmentfirst) -- (\tikzinputsegmentlast);
      },
    },
  },
  mid arrow/.style={postaction={decorate,decoration={
        markings,
        mark=at position .5 with {\arrow[#1]{stealth}}
      }}},
}

\newcommand{\ketB}[1]{\ket{#1}_B}
\newcommand{\ketC}[1]{\ket{#1}_C}
\newcommand{\ketD}[1]{\ket{#1}_D}

\newcommand{\braB}[1]{\ _{B\!}\bra{#1}}
\newcommand{\braC}[1]{\ _{C\!}\bra{#1}}
\newcommand{\braD}[1]{\ _{D\!}\bra{#1}}
\newcommand{\superp}[2]{\genfrac{}{}{0pt}{}{#1}{#2}}
\newcommand{\scalar}[2]{\langle #1|#2\rangle}
\newcommand{\scalarB}[2]{_{B}\langle #1|#2\rangle_{B}}

\newcommand{\scalarD}[2]{_{D}\langle #1|#2\rangle_{D}}

\newcommand{\dodd}[1]{\delta_{#1,\text{odd}}}
\newcommand{\deven}[1]{\delta_{#1,\text{even}}}
 
 \def\d{\delta}

 \def\p{\partial}
 
 \def\a{\alpha}
 \def\b{\beta}
 \def\g{\gamma}
 \def\d{\delta}
 \def\e{\varepsilon}
 
 \def\th{\theta}
 
 \def\k{\kappa}
 \def\l{\lambda}
 \def\m{\mu}

 \def\s{\sigma}
 
 \def\th{\theta}

 \def\G{\Gamma}
 \def\D{\Delta}

 \def\S{\Sigma}

 \def\o{\omega }

\def\CD{{\mathcal{D}}}

\def\CF{{\mathcal{F}}}

\def\CH{{\mathcal{H}}}

\def\CN{{\mathcal{N}}}

\def\CS{{\mathcal{S}}}
\def\CT{{\mathcal{T}}}

\def\CV{{\mathcal{V}}}
\def\CW{{\mathcal{W}}}

\def\hf{\dfrac{1}{2}}

\def\implies{\quad\Rightarrow\quad}

\def\tphi{\tilde{\phi}}

\def\trho{\tilde{\rho}}

\def\trho{\tilde{\rho}}

\def\CS{\mathcal{S}}

\def\hphi{\hat\phi}

\def\vac{\emptyset}

\def\mZ{\mathbb{Z}}
\def\mC{\mathbb{C}}

\def\bW{{\bar W}}

\def\gl{\mathfrak{gl}}

\def\glinf{\widehat{\mathfrak{gl}}(\infty)}
\def\oinf{\widehat{\mathfrak{o}}(\infty)}

\def\spinf{\widehat{\mathfrak{sp}}(\infty)}

\def\Abox{{\tikz[scale=0.007cm] \draw (0,0) rectangle (1,1);}}
\def\sAbox{{\tikz[scale=0.005cm] \draw (0,0) rectangle (1,1);}}

\def\Winf{W_{1+\infty}}

\def\tM{\tilde{M}}

\def\COC{\mathcal{O}^{(C)}}

\def\COB{\mathcal{O}^{(B)}}
\def\COBp{\mathcal{O}^{(B')}}
\def\COD{\mathcal{O}^{(D)}}

\def\hchi{{\hat \chi}}

\begin{document}
\preprint{CQUEST-2021-0657}

\title{Quantum $W_{1+\infty}$ subalgebras of BCD type and symmetric polynomials}
\author{Jean-Emile Bourgine}
\affiliation{Center for Quantum Spacetime (CQUeST), Sogang University, Seoul, 121-742, South Korea}%
\altaffiliation[Also at ]{Korea Institute for Advanced Studies (KIAS), Quantum Universe Center (QUC),85 Hoegiro, Dongdaemun-gu, Seoul, South Korea}
\email{bourgine@kias.re.kr}

\begin{abstract}
The infinite affine Lie algebras of type ABCD, also called $\glinf$, $\oinf$, $\spinf$, are equivalent to subalgebras of the quantum $W_{1+\infty}$ algebra. They have well-known representations on the Fock space of either a Dirac fermion ($\hat A_\infty$), a Majorana fermion ($\hat B_\infty$ and $\hat D_\infty$) or a symplectic boson ($\hat C_\infty$). Explicit formulas for the action of the quantum $W_{1+\infty}$ subalgebras on the Fock states are proposed for each representation. These formulas are the equivalent of the \textit{vertical presentation} of the quantum toroidal $\gl(1)$ algebra Fock representation. They provide an alternative to the fermionic and bosonic expressions of the \textit{horizontal presentation}. Furthermore, these algebras are known to have a deep connection with symmetric polynomials. The action of the quantum $W_{1+\infty}$ generators leads to the derivation of Pieri-like rules and q-difference equations for these polynomials. In the specific case of $\hat B_\infty$, a q-difference equation is obtained for $Q$-Schur polynomials indexed by strict partitions.
\end{abstract}

\maketitle

\section{Introduction}
The Lie algebra $\glinf$ is usually defined as a central extension of the algebra $\gl(\infty)$ of matrices with infinite size \cite{Kac1990}. Through its well-known representation on the Dirac fermion Fock space, it has been instrumental in finding the solutions of the Kadomtsev-Petviashvili (KP) hierarchy \cite{Date1981a}. Moreover, it also shares a deep relation with Schur symmetric polynomials that provide tau functions for the hierarchy. Its subalgebras of BCD type, namely $\hat B_\infty$, $\hat C_\infty$ and $\hat D_\infty$, have similar definitions, with further symmetry constraints imposed on the infinite matrices. They are naturally associated to the integrable hierarchies of type BKP, CKP and DKP \cite{Date1981,Jimbo1983}. The BKP and DKP hierarchies have polynomial tau functions given by the Q-Schur polynomials \cite{You1989,You1991}, a specialization of the Hall-Littlewood polynomials at $t=-1$ \cite{Schur1911,Macdonald}.\footnote{Instead, the specialization at $t=0$ and $t=1$ gives respectively the Schur polynomials and the symmetric monomials $m_\l$.} On the other hand, it was shown in \cite{Anguelova2017} that the CKP hierarchy does not admit any polynomial tau function. Yet, symmetric polynomials associated to the $\hat C_\infty$ algebra have been introduced recently by van de Leur, Orlov and Shiota in \cite{VanDeLeur2012}, and we will review their construction below. Most of these works focus on the important application to integrable hierarchies, but we would like to revisit here the relation between infinite Lie algebras and symmetric polynomials from a different perspective, namely quantum algebras.

As the name suggests, the quantum $W_{1+\infty}$ algebra is a q-deformation of the $W_{1+\infty}$ algebra \cite{KR,AFMO}. The algebra depends on a parameter $q\in\mC^\times$, and is defined in terms of the generators $W_{m,n}$, with indices $(m,n)\in\mZ\times\mZ\setminus\{(0,0)\}$, and a central element $C$, satisfying the commutation relations\footnote{When $n+n'=0$, the central term becomes $q^{-mn}mC\d_{m+m'}$.}
\begin{equation}\label{def_qW_C}
[W_{m,n},W_{m',n'}]=(q^{m'n}-q^{mn'})\left(W_{m+m',n+n'}+C\dfrac{\d_{m+m'}}{1-q^{n+n'}}\right).
\end{equation}
Here, we assume that $q$ is not a root of unity, and denote the algebra $\CW$ for short. This algebra can be realized in terms of $\glinf$ generators using a linear relation reminiscent of a discrete Fourier transform (see the equation \ref{rel_W_E} below). Subalgebras $\CW^X$ for $X=B,C,D$ can be defined as a quotient of $\CW$ by the group of automorphisms $\mZ_2=\{1,\s_X\}$ generated by the involutive automorphism $\s_X$ defined below. In the context of W-algebras, this procedure is called \textit{orbifolding} and we will use this terminology here. These subalgebras are realized in terms of the generators of the subalgebras $\hat X_\infty$ of $\glinf$ in the same way $\CW$ is \cite{Lebedev1992,Hoppe1993}.

The quantum toroidal algebra of $\gl(1)$, or Ding-Iohara-Miki algebra \cite{Ding1997,Miki2007}, is a refinement of the quantum $\Winf$ algebra.\footnote{The quantum toroidal algebra of $\gl(1)$ depends on two parameters $q_1,q_2$, in the limit $q_1q_2\to1$ it reduces to a version of the algebra $\CW$ with an extra central element $C'$ \cite{Bourgine2021}.} Following the inspirational paper \cite{AFS}, the former has played an important role in the context of topological string theory by providing an algebraic definition for the refined topological vertex \cite{Iqbal2007,Awata2008}. This algebraic framework led to enormous progress in the field, e.g. by extending the topological vertex technique to various theories \cite{BFMZ,Awata2017,Foda2018,Zenkevich2018,Bourgine2018,Kimura:2019gon,Bourgine:2019phm,Zenkevich2020} and observables \cite{BMZ,Bourgine2016,Bourgine2017b}, by deriving proofs for the q-deformed AGT correspondence \cite{Feigin2010,Awata2011,Fukuda2019}, or by describing the action of fiber-base duality \cite{Awata2009a,Bourgine2018a,SWM,Fukuda2019} to name only a few. It should be emphasized that the algebra $\CW$ plays the same role as the quantum toroidal $\gl(1)$ algebra in the self-dual limit of topological strings \cite{RTV1}, and that it can be used in the same way to build the original topological vertex \cite{Aganagic2003}, thus providing the quantum algebraic framework behind the melting crystal construction of Okounkov, Reshetikhin and Vafa \cite{ORV}.

Given the observed connections between topological strings and integrable hierarchies \cite{Aganagic2003,Nakatsu2007,Takasaki2018,Bourgine2021}, the relation between $\CW$ and $\glinf$ comes as no surprise. One of the motivation for this paper is to extend these connections to the subalgebras $\CW^X$. For this purpose, it appeared necessary to revisit first the relations between algebras and symmetric polynomials with a focus on the quantum $\Winf$ generators. To be specific, we examine the Fock representations of the algebras $\hat A_\infty,\hat B_\infty$, $\hat C_\infty$, $\hat D_\infty$, and derive explicit expressions for the action of the generators on the Fock states. We further associate to each algebra a set of symmetric polynomials indexed by the states defining the basis of the representation. Then, the action of the $\CW^X$-algebra induces identities among the polynomials: Pieri-like relations or q-difference equations.

Hence, we examine the level one representation of the $\glinf$ algebra on the Fock space of a 2D Dirac fermion. The states of the PBW basis are labeled by partitions, they are in one-to-one correspondence with the ring of symmetric polynomials in infinitely many variables. We derive explicit combinatorial formulas for the action of the generators $W_{m,n}$ on the states associated to Schur polynomials. From these expressions, we recover the rules obeyed by the Schur polynomials under multiplication by elementary power sums, together with the q-difference equation coming from the specialization of the Macdonald operator at $t=q$ \cite{Macdonald}. Expanding this operator in $q$, we recover the Hamiltonians of the Calogero-Sutherland integrable model \cite{Sutherland1988}. This program is then extended to the subalgebras $\hat B_\infty$, $\hat C_\infty$ and $\hat D_\infty$ with a mitigated success. In the case of $\hat B_\infty$, we study the representation of level $1/2$ on the Fock space of a Majorana fermion. As anticipated, the associated symmetric polynomials are $Q$-Schur functions, and the whole program goes through. In particular, we obtain a new q-difference equation solved by these polynomials. On the other hand, the definition of symmetric polynomials is not obvious in the case of the algebras $\hat C_\infty$ and $\hat D_\infty$. The corresponding representations have levels $C=-1/2$ and $C=1/2$ respectively, they are realized on the Fock space of a symplectic boson or a Majorana fermion. For these algebras, there are two natural ways to generalize the definition of Schur and Q-Schur functions of the cases $\hat A_\infty$ and $\hat B_\infty$, and we study the properties of both sets of symmetric polynomials.

The organization of this paper is very straightforward, the sections two to five treating successively the case of the algebras $\hat A_\infty,\hat B_\infty$, $\hat C_\infty$ and $\hat D_\infty$. The first appendix re-derive the bosonic expression of the Schur polynomials, while the second appendix focuses on the derivation of the q-difference equations.

\begin{figure}
\begin{center}
\begin{tikzpicture}[scale=0.7]
\draw [fill=white] (1,1) rectangle (2,2);
\draw [fill=white] (1,2) rectangle (2,3);
\draw [fill=white] (1,3) rectangle (2,4);
\draw [fill=white] (1,4) rectangle (2,5);
\draw [fill=white] (2,1) rectangle (3,2);
\draw [fill=white] (2,2) rectangle (3,3);
\draw [fill=white] (2,3) rectangle (3,4);
\draw [fill=white] (3,1) rectangle (4,2);
\draw [fill=white] (3,2) rectangle (4,3);
\draw [fill=white] (3,3) rectangle (4,4);
\draw [fill=white] (4,1) rectangle (5,2);
\draw [fill=white] (5,1) rectangle (6,2);
\end{tikzpicture}
\end{center}
\caption{Young diagram of the partition $\l=(4,3,3,1,1)$ with three hooks of length $8,3,1$}
\label{fig1}
\end{figure}
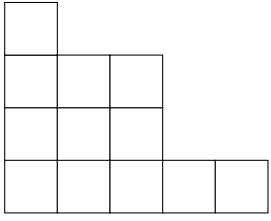

\para{Notations} In this paper, we take the convention that indices $i,j,k,l,m,n$ take integer values, while indices $r,s,t$ take half-integer values. Moreover, partitions are denoted with the Greek letters $\l,\mu,\nu$, they consists in sets of ordered positive integers $\l=(\l_1\geq \l_2\geq\cdots)$ with $|\l|=\sum_i\l_i$ finite. They are represented as Young diagrams with columns of $\l_i$ boxes, drawn upward (see the figure \ref{fig1}). We denote $\ell(\l)=\sharp\{\l_i\neq0\}$ the number of parts, $|\l|$ the number of boxes and $d(\l)$ the number of hooks.

\section{Quantum $\Winf$ and $\glinf$}\label{sec_def_rho}
The algebra $\glinf$, or $\hat A_\infty$, can be formulated in terms of the generators $E_{r,s}$ with half-integer indices, and a central element $C$, obeying the commutation relations,
\begin{align}
\begin{split}\label{Ers}
&\left[E_{r,s}, E_{t,u}\right]=\delta_{s+t}E_{r,u}-\delta_{r+u}E_{t,s} +C(\theta(r)-\theta(t))\delta_{s+t}\delta_{r+u},\\
\end{split}
\end{align}
with $r,s,t,u\in \mathbb{Z}+1/2$ and where $\theta(r)$ denotes the Heaviside function taking the value one for $r>0$ and zero when $r<0$ (the value at zero is irrelevant at the moment). By construction, it has a representation of level zero given by infinite matrices, namely $\rho_{0}(E_{r,s})=e_{r+1/2,-s+1/2}$ where $e_{i,j}$ ($i,j\in\mZ$) are the matrices with $1$ at position $(i,j)$ and zero elsewhere that form a basis of $\mathfrak{gl}(\infty)$.

The algebraic relations \ref{def_qW_C} defining the quantum $\Winf$ algebra $\CW$ can be realized in terms of the generators $E_{r,s}$ of $\glinf$ after introducing the linear relation
\begin{equation}\label{rel_W_E}
W_{m,n}=\sum_{r\in\mZ+1/2}q^{-(r+1/2)n}E_{m-r,r},
\end{equation}
and upon identification of the central elements $C$ \cite{Lebedev1992}. Through this relation, the representations of $\glinf$ are naturally lifted to representations of quantum $\Winf$. In this paper, we will consider only two representations referred as \textit{Dirac representation} and \textit{$\b\g$-representation}.

As the name suggests, the Dirac representation $\rho^{(D)}$ acts on the Fock space of a 2D Dirac fermion with modes $\psi_r$, $\bpsi_r$ satisfying the Clifford algebra $\{\psi_r,\bpsi_s\}=\d_{r+s}$. The Fock space is built upon the vacuum state $\ket{\vac}$, annihilated by positive modes, from the action of negative modes. The representation has level one and follows from the assignment $\rho^{(D)}(E_{r,s}) = :\bpsi_r\psi_s:$ where $::$ denotes the normal ordering with positive modes moved to the right \cite{Frenkel1994}. It is deeply connected to the Schur polynomials and will be studied in great details in this section.

On the other hand, the $\b\g$-representation $\rho^{(\b\g)}$ has level minus one. It acts on the Fock space of the $(\b,\g)$ bosonic ghost system with modes $\b_r$, $\g_r$ satisfying the Heisenberg algebra $[\b_r,\g_s]=-\d_{r+s}$ \cite{KR1995}. The states are built in the same manner from a vacuum state $\ket{\vac}$ annihilated by the positive modes of both fields, $\beta_r\ket{\vac}=\gamma_r\ket{\vac} =0$ for $r>0$, and the generators are also represented in the same way $\rho^{(\b\g)}(E_{r,s}) = :\b_r\g_s:$. 

Both algebras $\glinf$ and $\CW$ admit a co-commutative coproduct, and representations of higher integer level $|C|>1$ are easily obtained by taking the tensor product of $|C|$ Dirac or $\b\g$ representations.

\para{Remark} In the limit $q\to1$, the algebra $\CW$ reduces to the $W_{1+\infty}$ algebra and the modes $W_{m,n}$ can be expanded in powers of $(n\a)$ where $q=e^{-\a}$. The $k$th power $W_m^{(k)}$ corresponds to a linear combination of the $m$th modes of the currents of spin lower or equal to $k+1$,
\begin{equation}
W_{m,n}=\sum_{k=0}^{\infty}\dfrac{(n\a)^k}{k!}W_m^{(k)},\quad W_m^{(k)}=\sum_{r\in\mZ+1/2}(r+1/2)^k E_{m-r,r}.
\end{equation} 
Most of the results presented in this paper have their counterpart in the $W_{1+\infty}$ algebra. In particular, the commutative subalgebra appearing in the vertical presentation coincides with the zero modes of the currents of spin $k$ that are indeed commutative. For more information on the relations between BCD subalgebras of $\glinf$ and subalgebras of $W_{1+\infty}$, we refer to the very complete paper of Kac, Wang and Yan \cite{Kac1998}. Their results may be relevant to the study of the $4D$ $\CN=2$ gauge theories in the self-dual omega-background (i.e. $\mC_{\e_1}\times\mC_{\e_2}$ at $\b=-\e_1/\e_2=1$) \cite{Bourgine2018}.

\subsection{Dirac representation}\label{sec_Dirac}
We denoted $\rho^{(D)}$ the Dirac representation of the algebra $\CW$, but omit the notation if no confusion ensues. The representation of the generators $W_{m,n}$ follows from the relation \ref{rel_W_E},
\begin{equation}\label{W_psi}
W_{m,n}=\sum_{r\in\mZ+1/2}q^{-(r+1/2)n}:\bpsi_{m-r}\psi_r:.
\end{equation} 
Introducing the fermionic fields
\begin{equation}
\psi(z)=\sum_{r\in\mathbb{Z}+1/2}z^{-r-1/2}\psi_r,\quad \bpsi(z)=\sum_{r\in\mathbb{Z}+1/2}z^{-r-1/2}\bpsi_r
\end{equation}
that satisfy the anticommutation relation $\{\psi(z),\bpsi(w)\}=z^{-1}\d(z/w)$ with the multiplicative Dirac delta function $\d(z)=\sum_{k\in\mZ}z^k$, the generators can be realized as the contour integrals
\begin{equation}\label{W_contour}
W_{m,n}=\oint{\dfrac{dz}{2i\pi}z^m:\bpsi(z)\psi(q^nz):}.
\end{equation}
The adjoint action of the algebra $\CW$ on the fermionic fields reads
\begin{align}
\begin{split}
&[W_{m,n},\psi(z)]=-z^m\psi(q^nz),\\
&[W_{m,n},\bpsi(z)]=q^{(m-1)n}z^m\bpsi(q^{-n}z).
\end{split}
\end{align}

The Dirac fermion Fock space has a PBW basis spanned by the states $\ket{\l}$ labeled by Young diagrams $\l=(\l_1,\l_2,\cdots)$,
\begin{equation}\label{def_Schur}
\ket{\l}=\prod_{i=1}^{d(\l)}\left[(-1)^{\l'_i+i}\bar\psi_{-(\l_i-i+1/2)}\psi_{-(\l'_i-i+1/2)}\right]\ket{\vac},
\end{equation}
where $\l'$ is the transposed of the Young diagram $\l$, and $d(\l)$ the number of boxes on the diagonal. We recognize in the indices the arm length $a(\Abox)=\l_i-j$ and leg length $l(\Abox)=\l'_j-i$ of the boxes $\Abox=(i,j)\in\l$ on the diagonal (i.e. $i=j$). These states are called \textit{Schur states} for a reason that will become obvious shortly.

\para{Bosonization} The Dirac representation can be rewritten as an action on the Fock space of a 2D free boson using the celebrated bosonization technique. The bosonic Fock space is built upon the action of the negative modes $a_{-k}$ of an Heisenberg algebra $[a_k,a_l]=k\d_{k+l}$ on the vacuum $\ket{\vac}$ annihilated by positive modes. The bosonization formulas,
\begin{align}\label{exp_phi}
\begin{split}
&\bar\psi(z)=\vdots e^{\phi(z)}\vdots,\quad  \psi(z)=\vdots e^{-\phi(z)}\vdots,\quad \vdots\bpsi(z)\psi(z)\vdots=\p\phi(z),\\
&\text{with}\quad \phi(z)=Q+a_0\log z-\sum_{k\in\mZ^\times}\dfrac1kz^{-k}a_k,
\end{split}
\end{align}
also involve a zero mode $a_0$ and its dual operator $Q$ such that $[a_0,Q]=1$. To distinguish it from the fermionic normal-ordering, we denote the bosonic normal-ordering with the symbols $\vdots\cdots\vdots$, it consists in moving to the right the modes $a_k$ with $k\geq0$. Strictly speaking, the fermionic Fock space $\CF$ decomposes into a direct sum of bosonic modules $\CV_{\a}$ in which $a_0$ acts as a constant $\a$. However, the representation of the quantum $\Winf$ generators do not involve the operator $Q$, thus $a_0$ is central and we can restrict ourselves to a single module $\CV_\a$, we take here $\a=0$.

The relation between bosonic and fermionic modes
\begin{equation}\label{ak_psi}
a_k=\sum_{r\in\mZ+1/2} :\bar\psi_{k-r}\psi_r:
\end{equation} 
leads to identify the Heisenberg modes $a_k$ with the Dirac representation of the generators $W_{k,0}$ of $\CW$. The representation for the modes $W_{m,n}$ with $n\neq0$ is more easily formulated using the currents
\begin{equation}\label{def_wn}
w_n(z)=(1-q^n)\sum_{m\in\mZ}z^{-m}W_{m,n}+C,
\end{equation} 
for which the representation reads
\begin{align}
\begin{split}\label{wn_boson}
w_n(z)&=(1-q^n)z\bpsi(z)\psi(q^nz)\\
&=q^{-na_0}\vdots\exp\left(-\sum_{k\in\mZ^\times}\dfrac{z^{-k}}{k}(1-q^{-nk})a_k\right)\vdots.
\end{split}
\end{align}

The modules $\CV_\a$ are isomorphic to the ring of symmetric polynomials with infinitely many variables. This correspondence sends the symmetric power sums $p_k(x)=\sum_ix_i^k$ to the creation operators $a_{-k}$, and the constant polynomial $1$ to the vacuum state $\ket{\vac}$. Moreover, it follows from the Frobenius formula that the Schur states $\ket{\l}$ defined in \ref{def_Schur} are the image of the Schur polynomials $s_\l(x)$ \cite{Marino}. They can be computed explicitly by decomposition of these polynomials in the PBW basis of symmetric power sums, replacing 
\begin{equation}
p_{\l_1}(x)^{k_1}\cdots p_{\l_n}(x)^{k_n} \leftrightarrow (a_{-\l_1})^{k_1}\cdots (a_{-\l_n})^{k_n}\ket{\vac}.
\end{equation} 
It is also possible to define the adjoint basis acting with $a_k=(a_{-k})^\dagger$ and $Q^\dagger=-Q$, $a_0^\dagger=a_0$ (or $\bpsi_r^\dagger=\psi_{-r}$, $\psi_r^\dagger=\bpsi_{-r}$) on the dual vacuum $\bra{\vac}$. The Schur states obtained in this way are orthonormal, $\scalar{\l}{\mu}=\d_{\l,\mu}$.

\subsection{Vertical presentation}
The quantum $\Winf$ algebra is a limit of the quantum toroidal $\gl(1)$ algebra, or Ding-Iohara-Miki algebra \cite{Ding1997,Miki2007}. The latter is known to possess two central elements $(c_1,c_2)$, and two associated Heisenberg subalgebras $[a_k^{(i)},a_l^{(i)}]=c_ik\d_{k+l}$ (upon normalization). The Fock representation has levels $(1,0)$, so that only one of the two Heisenberg subalgebra survives, the other one producing an infinite set of commuting operators. This property leads to two different \textit{presentations} of the Fock representation. In the \textit{horizontal presentation}, the focus is on the Heisenberg algebra $a_k^{(1)}$ with respect to which the generators are expressed as vertex operators \cite{Feigin2009a}. On the other hand, in the \textit{vertical presentation}, the focus is on the commuting operators $a_k^{(2)}$ that act diagonally on a chosen basis \cite{feigin2011quantum}. These two dual pictures are exchanged by the action of Miki's automorphism that switches the two subalgebras and maps $(c_1,c_2)\to(-c_2,c_1)$ \cite{Miki2007}.\footnote{We insist here on the term \textit{presentation} because in this picture the representation $\rho$ is fixed, we simply express the action of the generators differently: the horizontal presentation refers to the expression of the Drinfeld currents $x^\pm(x)$, $\psi^\pm(z)$ as vertex operators while the vertical presentation refers to the action of the dual currents $\CS\cdot x^\pm(z)$, $\CS\cdot\psi^\pm(z)$ on the Fock states ($\CS$ denoting Miki's automorphism).} 

The quantum $\Winf$ algebra inherits of the same properties. In the definition \ref{def_qW_C}, the second central charge $c_2$ has already been set to zero, while we can identify $c_1=C$ \cite{Bourgine2021}. The two subalgebras correspond to $a_k^{(1)}=W_{k,0}$ and $a_k^{(2)}=W_{0,k}$. The horizontal presentation can be read from the equation \ref{wn_boson} that indeed expresses the currents $w_n(z)$ built upon the generators $W_{m,n}$ as vertex operators in the modes $a_k^{(1)}=a_k$. In this subsection, we derive the vertical presentation in which the action of $a_k^{(2)}=W_{0,k}$ is diagonal. Roughly speaking, Miki's automorphism sends $W_{m,n}$ to $W_{-n,m}$, thereby exchanging the two subalgebras \cite{SWM,Bourgine2021}.

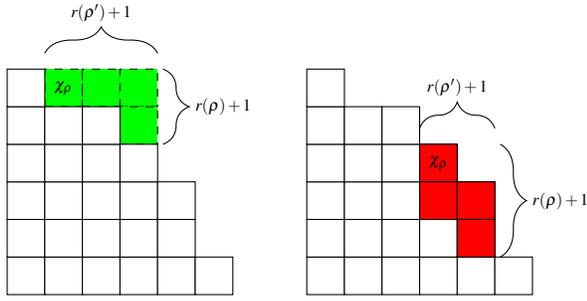
\begin{figure}
\begin{center}
\begin{tikzpicture}[scale=0.5]
\draw [fill=green,dashed] (2,6) rectangle (3,7);
\draw [fill=green,dashed] (3,6) rectangle (4,7);
\draw [fill=green,dashed] (4,6) rectangle (5,7);
\draw [fill=green,dashed] (4,5) rectangle (5,6);
\node[scale=.7] at (2.5,6.5) {$\chi_\rho$};
\draw [decorate,decoration={brace,amplitude=10pt,mirror,raise=4pt},xshift=-4pt,yshift=0pt] (5,5) -- (5,7) node [black,midway,right,xshift=14pt,scale=.7]  {$r(\rho)+1$};
\draw [decorate,decoration={brace,amplitude=10pt,mirror,raise=4pt},xshift=0pt,yshift=4pt] (5,7) -- (2,7) node [black,midway,above,yshift=14pt,scale=.7]  {$r(\rho')+1$};
\draw [fill=white] (1,1) rectangle (2,2);
\draw [fill=white] (1,2) rectangle (2,3);
\draw [fill=white] (1,3) rectangle (2,4);
\draw [fill=white] (1,4) rectangle (2,5);
\draw [fill=white] (1,5) rectangle (2,6);
\draw [fill=white] (1,6) rectangle (2,7);
\draw [fill=white] (2,1) rectangle (3,2);
\draw [fill=white] (2,2) rectangle (3,3);
\draw [fill=white] (2,3) rectangle (3,4);
\draw [fill=white] (2,4) rectangle (3,5);
\draw [fill=white] (2,5) rectangle (3,6);
\draw [fill=white] (3,1) rectangle (4,2);
\draw [fill=white] (3,2) rectangle (4,3);
\draw [fill=white] (3,3) rectangle (4,4);
\draw [fill=white] (3,4) rectangle (4,5);
\draw [fill=white] (3,5) rectangle (4,6);
\draw [fill=white] (4,1) rectangle (5,2);
\draw [fill=white] (4,2) rectangle (5,3);
\draw [fill=white] (4,3) rectangle (5,4);
\draw [fill=white] (4,4) rectangle (5,5);
\draw [fill=white] (5,1) rectangle (6,2);
\draw [fill=white] (5,2) rectangle (6,3);
\draw [fill=white] (5,3) rectangle (6,4);
\draw [fill=white] (6,1) rectangle (7,2);
\end{tikzpicture}
\hspace{5mm}
\begin{tikzpicture}[scale=0.5]
\draw [fill=white] (1,1) rectangle (2,2);
\draw [fill=white] (1,2) rectangle (2,3);
\draw [fill=white] (1,3) rectangle (2,4);
\draw [fill=white] (1,4) rectangle (2,5);
\draw [fill=white] (1,5) rectangle (2,6);
\draw [fill=white] (1,6) rectangle (2,7);
\draw [fill=white] (2,1) rectangle (3,2);
\draw [fill=white] (2,2) rectangle (3,3);
\draw [fill=white] (2,3) rectangle (3,4);
\draw [fill=white] (2,4) rectangle (3,5);
\draw [fill=white] (2,5) rectangle (3,6);
\draw [fill=white] (3,1) rectangle (4,2);
\draw [fill=white] (3,2) rectangle (4,3);
\draw [fill=white] (3,3) rectangle (4,4);
\draw [fill=white] (3,4) rectangle (4,5);
\draw [fill=white] (3,5) rectangle (4,6);
\draw [fill=white] (4,1) rectangle (5,2);
\draw [fill=white] (4,2) rectangle (5,3);
\draw [fill=white] (4,3) rectangle (5,4);
\draw [fill=white] (4,4) rectangle (5,5);
\draw [fill=white] (5,1) rectangle (6,2);
\draw [fill=white] (5,2) rectangle (6,3);
\draw [fill=white] (5,3) rectangle (6,4);
\draw [fill=white] (6,1) rectangle (7,2);
\draw [fill=red] (4,4) rectangle (5,5);
\draw [fill=red] (4,3) rectangle (5,4);
\draw [fill=red] (5,3) rectangle (6,4);
\draw [fill=red] (5,2) rectangle (6,3);
\node[scale=.7] at (4.5,4.5) {$\chi_\rho$};
\draw [decorate,decoration={brace,amplitude=10pt,mirror,raise=4pt},xshift=-4pt,yshift=0pt] (6,2) -- (6,5) node [black,midway,right,xshift=14pt,scale=.7]  {$r(\rho)+1$};
\draw [decorate,decoration={brace,amplitude=10pt,mirror,raise=4pt},xshift=0pt,yshift=4pt] (6,5) -- (4,5) node [black,midway,above,yshift=14pt,scale=.7]  {$r(\rho')+1$};
\end{tikzpicture}
\end{center}
\caption{Example of the Young diagram $65^2431$ and a border strip $\rho$ with either $\rho\in A_4(\l)$ (left) or $\rho\in R_4(\l)$ (right)}
\label{fig2}
\end{figure}

It is not difficult to derive the vertical presentation from the fermionic formulas for the states \ref{def_Schur} and the generators \ref{W_psi}. However, in order to write down this action in a form similar to the vertical presentation of the quantum toroidal $\gl(1)$ algebra in \cite{Bourgine2017b,Bourgine2018a}, it is necessary to introduce several notations. Following ref. \cite{Macdonald}, we call a border strip of length $m$ a set of $m$ adjacent boxes (i.e. sharing a single edge). We denote $A_m(\l)$ (resp. $R_m(\l)$) the set of border strips of lengths $m$ that can be added to (resp. removed from) the Young diagram $\l$ such that $\l\pm\rho$ is a valid Young diagram. In addition, to any box $\Abox$ of a Young diagram, we assign a complex number called the \textit{box content}, and obtained as $\chi_{\sAbox}=q^{i-j}\in\mC^\times$ where $(i,j)$ are the coordinates of the box. For each strip $\rho\in A_m(\l)\cup R_m(\l)$, we associate the complex number $\chi_\rho$ corresponding to the content of the top left box. Thus, the contents of the boxes in the strip $\rho$ are given by $q^{i-1}\chi_\rho$ for $i=1\cdots m$.\footnote{Note that the profile of the Young diagram meet the diagonals $i-j=$fixed at only one point. As a result, the content $\chi_\rho$ characterizes uniquely the strips $\rho\in A_m(\l)\cup R_m(\l)$ at fixed $m$.} Finally, we denote $r(\rho)$ (resp. $r(\rho')$) the number of rows (columns) that a strip $\rho$ occupies minus one, we have $r(\rho)+r(\rho')=m-1$ (see figure \ref{fig2}).

With these notations, the action on the Schur states read ($m>0$)
\begin{align}
\begin{split}\label{Wmn_Fock}
&W_{0,n}\ket{\l}=-(1-q^{-n})\sum_{\sAbox\in\l}\chi_{\sAbox}^n\ \ket{\l},\\
&W_{m,n}\ket{\l}=q^{-n}\sum_{\rho\in R_m(\l)}(-1)^{r(\rho')}\chi_\rho^n\ket{\l-\rho},\\
&W_{-m,n}\ket{\l}=q^{(m-1)n}\sum_{\rho\in A_m(\l)}(-1)^{r(\rho')}\chi_\rho^n\ket{\l+\rho},
\end{split}
\end{align}
and, in particular,
\begin{align}
\begin{split}\label{action_ak}
&a_k\ket{\l}=\sum_{\rho\in R_k(\l)}(-1)^{r(\rho')}\ket{\l-\rho},\\
&a_{-k}\ket{\l}=\sum_{\rho\in A_k(\l)}(-1)^{r(\rho')}\ket{\l+\rho}.
\end{split}
\end{align}
The algebra can be supplemented with a grading operator $d$ such that $[d,W_{m,n}]=-mW_{m,n}$. It is represented as
\begin{align}
\begin{split}\label{def_d}
&d=\sum_{r\in\mZ+1/2}r:\bpsi_{-r}\psi_r:\\
\implies &d\ket{\l}=|\l|\ket{\l},\quad \a^dW_{m,n}\a^{-d}=\a^{-m}W_{m,n}.
\end{split}
\end{align}

\subsection{Symmetric polynomials}\label{sec_Schur}
The correspondence between the Schur states $\ket{\l}$ of the bosonic Fock space and the Schur symmetric polynomials $s_\l(x)$ provides an expression for the Schur polynomials as bosonic correlators,
\begin{equation}\label{Schur_corr}
s_\l(x)=\bra{\vac}e^{\sum_{k>0}\frac{p_k(x)}{k}a_{k}}\ket{\l}.
\end{equation}
In practice, the polynomial is obtained by expanding the exponential as an infinite series. Since $a_k$ acts on the Schur states by removing strips of boxes (following equ. \ref{action_ak}), the r.h.s. contains only a finite number of terms. The proof of this formula is not easily found in the litterature and we decided to include a short derivation in the appendix \ref{AppA}. 

As we shall see, the formula \ref{Schur_corr} for the Schur polynomials, combined with the expression \ref{action_ak} for the vertical action of the generators $W_{m,n}$, can be used to recover several important properties of the Schur polynomials. As a warm-up, we can introduce the exponentiated grading operator $\a^d$ on the left of the correlator since $\bra{\vac}\a^d=\bra{\vac}$. Commuting the operator all the way to the right using the formulas \ref{def_d}, we recover the identity $s_\l(\a x)=\a^{|\l|}s_\l(x)$ expressing the fact that Schur polynomials are homogeneous of degree $|\l|$.

Recursion relations, reminiscent of the Pieri rules, are obtained in a similar manner from the insertion of the operator $a_{-k}$ inside the correlator \ref{Schur_corr}. Commuting the operator with the exponential, we find
\begin{equation}
\bra{\vac}a_{-k}e^{\sum_{k>0}\frac{p_k(x)}{k}a_{k}}\ket{\l}=-p_k(x)s_\l(x)+\bra{\vac}e^{\sum_{k>0}\frac{p_k(x)}{k}a_{k}}a_{-k}\ket{\l}.
\end{equation}
Evaluating the action of $a_{-k}$ on the Schur states using the vertical presentation \ref{action_ak}, we recover indeed the rules (see \cite{Macdonald}, page 48)
\begin{equation}
p_k(x)s_\l(x)=\sum_{\rho\in A_k(\l)}(-1)^{r(\rho')}s_{\l+\rho}(x).
\end{equation}
The usual Pieri rules for Schur polynomials involve either the complete symmetric functions $h_k$ or the elementary symmetric function $e_k$. Here, instead, we multiply by the elementary power sums $p_k$, which explains the presence of the extra signs in the r.h.s..

This derivation can be easily generalized to skew-Schur polynomials using the formula \ref{slm_Gamma} of the appendix and the contragredient action of $a_{-k}$, we find
\begin{align}
\begin{split}
p_k(x)s_{\l/\mu}(x)=&\sum_{\rho\in A_k(\l)}(-1)^{r(\rho')}s_{(\l+\rho)/\mu}(x)\\
&-\sum_{\rho\in R_k(\mu)}(-1)^{r(\rho')}s_{\l/(\mu-\rho)}(x).
\end{split}
\end{align}
Furthermore, taking the derivative $\p/\p p_k$ on both sides of the equation \ref{Schur_corr}, and using the expression \ref{action_ak} for the action of $a_k$ on Schur states, we find
\begin{equation}
k\dfrac{\p}{\p p_k} s_\l(x)=\sum_{\rho\in R_k(\l)}(-1)^{r(\rho')}s_{\l-\rho}(x).
\end{equation}

The algebra $\CW$ is multiplicatively generated by the modes $W_{m,0}$ and $W_{0,n}$, so it is sufficient to examine the action of the remaining modes $W_{0,n}$. As shown in appendix \ref{AppB}, the insertion of these modes in the correlator \ref{Schur_corr} generates a q-difference equation solved by the Schur polynomials, namely
\begin{align}
\begin{split}\label{qdiff_Schur}
&\sum_{i=1}^N\prod_{j\neq i}\dfrac{x_i-q^nx_j}{x_i-x_j}T_{q^{-n},x_i} s_\l(x)\\
=&\left(\sum_{i=1}^N q^{n(i-1)}+q^{-n}(1-q^{n})\sum_{\sAbox\in\l}\chi_{\sAbox}^n\right)s_\l(x),
\end{split}
\end{align}
where, following \cite{Macdonald}, we denoted $T_{q^{-n},x_i}$ the operation consisting in replacing the variable $x_i$ by $q^{-n}x_i$. The number $N>\ell(\l)$ of variables $x_i$ has been re-introduced to play the role of a cut-off. In this equation, the dependence in $n$ arises only through the variable $q^n$, and thus we have only a single q-difference equation. 
In fact, the l.h.s. can be identified with the Macdonald operator $t^{1-N}D_N^1$ after the replacement $(t,q)\to (q^{-n},q^{-n})$ (recall that Schur polynomials are a specialization of Macdonald polynomials at $t=q$). The dependence in the number $N$ of variables drops if we define
\begin{align}
\begin{split}\label{def_E}
&E=q^n\sum_{i=1}^N\prod_{j\neq i}\dfrac{x_i-q^nx_j}{x_i-x_j}T_{q^{-n},x_i}-\sum_{i=1}^N q^{ni}\\
\implies &E s_\l(x)=(1-q^n)\sum_{\sAbox\in\l}\chi_{\sAbox}^n\ s_\l(x).
\end{split}
\end{align}
This expression is suitable for taking the limit $N\to\infty$.

\para{Macdonald polynomials} The presence of the Heisenberg subalgebra formed by the modes $W_{k,0}$ is essential for the derivation of these formulas. The quantum toroidal $\gl(1)$ algebra also contains a Heisenberg subalgebra and the same considerations apply to Macdonald polynomials. The derivation of the appendix \ref{AppA} extends to this case, and provides the formula
\begin{equation}
P_\l(x)=\bra{\vac}e^{\sum_{k>0}\frac{p_k(x)}{k}\frac{1-t^k}{1-q^k}a_k}\ket{P_\l},
\end{equation}
in terms of the Macdonald states $\ket{P_\l}$ \cite{Bourgine2018a}. Then, the combination of horizontal and vertical actions produce both the Pieri rules and the Macdonald operators \cite{Feigin2009a}. In the next sections, we will attempt to apply this generic program to the subalgebras $\hat B_\infty$, $\hat C_\infty$ and $\hat D_\infty$.

\para{Hamiltonians} We would like to conclude this section with an important remark. Since the Schur polynomials do not depend on the parameter $q^n$, the q-difference equation \ref{qdiff_Schur} produces in the limit $q\to1$ an infinite tower of differential equations all satisfied by the Schur polynomials. Setting $q^n=e^{-\a}$, we can derive easily the first two differential operators in the limit $\a\to 0$,\footnote{The expansion is easily done after conjugation by the Van der Monde determinant $\D(x)=\prod_{i<j}(x_j-x_i)$ since
\begin{equation}
\sum_{i=1}^N\prod_{j\neq i}\dfrac{q^{-n}x_i-x_j}{x_i-x_j}T_{q^{-n},x_i}=\D(x)^{-1}\sum_iT_{q^{-n},x_i} \D(x).
\end{equation}\protect\trick.}
\begin{equation}
\sum_{i=1}^N\prod_{j\neq i}\dfrac{q^{-n}x_i-x_j}{x_i-x_j}T_{q^{-n},x_i}=N+\a\CH_1+\a^2\CH_2+O(\a^2).
\end{equation} 
The Hamiltonian $\CH_k$ gives the action of the zero mode of the current of spin $k$ for the $W_{1+\infty}$ algebra found in the degenerate limit $q\to1$. After a short calculation, we find the expressions
\begin{align}
\begin{split}
&\CH_1=\sum_iD_i+\hf N(N-1),\\
&\CH_2=\hf\CH_\text{CS}^{\b=1}+\hf(N-1)\CH_1-\dfrac1{12}N(N-1)(N-2),
\end{split}
\end{align} 
where we denoted $D_i=x_i\p_{x_i}$ and $\CH_\text{CS}^{\b=1}$ is the Calogero-Sutherland Hamiltonian specialized at $\b=1$,
\begin{equation}
\CH_\text{CS}^{\b}=\sum_iD_i^2+\b\sum_{\superp{i,j}{i<j}}\dfrac{x_i+x_j}{x_i-x_j}(D_i-D_j).
\end{equation} 
Expanding also the r.h.s. of the q-difference equation \ref{qdiff_Schur}, we find $\sum_iD_i s_\l(x)= |\l| s_\l(x)$ and
\begin{align}
\begin{split}
&\CH_\text{CS}^{\b=1}s_\l(x)=\left(\k(\l)+N|\l|\right)s_\l(x),\\
\text{with}\quad &\k(\l)=2\sum_{(i,j)\in\l}(j-i).
\end{split}
\end{align}
The first equation expresses the homogeneity of the Schur polynomials, while the second equation states that they are eigenfunctions of the Calogero-Sutherland Hamiltonian at $\b=1$. Of course, this fact is not a surprise since Jack polynomials are eigenfunctions of the Calogero-Sutherland Hamiltonian for generic $\b$, and they specialize to Schur polynomials at $\b=1$.
%

\section{$B$-type orbifold}
The algebra $\hat B_\infty$ is a central extension of the algebra of infinite matrices $b_{i,j}$ with integer indices that satisfy the relation $b_{i,j}=(-1)^{i+j+1}b_{-j,-i}$ \cite{Kac1990}. It can be defined using the generators $\COB_{r,s}$ with half-integer indices and the central element $C$ satisfying the commutation relations
\begin{align}
\begin{split}
[\COB_{r,s},\COB_{t,u}]&=\d_{s+t}\COB_{r,u}-\d_{r+u}\COB_{t,s}+2C\d_{s+t}\d_{r+u}(\th(r)-\th(t))\\
&+(-1)^{t+u}\Big[\d_{r+t-1}\COB_{u+1,s}-\d_{s+u+1}\COB_{r,t-1}\\
&\quad+2C\d_{s+u+1}\d_{r+t-1}(\th(u+1)-\th(r))\Big],
\end{split}
\end{align}
together with the property $\COB_{r,s}+(-1)^{r+s}\COB_{s+1,r-1}=-2C\d_{r+s}\d_{s+1/2}$. In the r.h.s. of the commutator, the first line is identical to the relation \ref{Ers} for the generators of $\glinf$ while the second line is new. This is a recurrent feature of the subalgebras of $\glinf$ considered in this paper, it is due to the fact that $\hat B_\infty$ is a subalgebra of $\glinf$ generated by the linear combinations
\begin{equation}\label{def_COB}
\COB_{r,s}=E_{r,s}+(-1)^{r+s+1}E_{s+1,r-1}-C\d_{r+s}\d_{s+1/2}.
\end{equation} 

Applying the linear transformation \ref{rel_W_E} to the operators $\COB_{r,s}$, we find the generators
\begin{equation}\label{def_WB}
W^B_{m,n}=\sum_{r\in\mZ+1/2}q^{-(r+1/2)n}\COB_{m-r,r}
\end{equation} 
of a subalgebra of $\CW$ that we denote $\CW^B$. It was shown in \cite{Lebedev1992} that this subalgebra can be realized as a quotient of $\CW$ by the group $\mZ_2$ of automorphisms generated by an involution $\s_B$ acting on the generators as follows
\begin{equation}\label{def_sB}
\s_B:\ W_{m,n}\to (-1)^{m+1}q^{-mn}W_{m,-n}-C\d_{m,0},\quad C\to C.
\end{equation} 
Thus, the subalgebra $\CW^B$ is generated by the central element $(1+\s_B)C=2C$ and the symmetric projections $W^B_{m,n}=(1+\s_B)W_{m,n}$ that coincide with the definitions \ref{def_WB}. From the action of $\s_B$, we deduce the commutation relations satisfied by the generators $W_{m,n}^B$,
\begin{align}
\begin{split}\label{com_WB}
&[W_{m,n}^B,W_{m',n'}^B]=(q^{m'n}-q^{mn'})\left(W^B_{m+m',n+n'}+2C\dfrac{\d_{m+m'}}{1-q^{n+n'}}\right)+\\
&+(-1)^{m'+1}q^{-m'n'}(q^{m'n}-q^{-mn'})\left(W^B_{m+m',n-n'}+2C\dfrac{\d_{m+m'}}{1-q^{n-n'}}\right).
\end{split}
\end{align}
Note that the modes $W_{m,0}$ with $m$ even are projected out, while the odd modes define the Heisenberg subalgebra $[W_{m,0}^B,W_{m',0}^B]=4Cm\d_{m+m'}$. They will be used to define the horizontal presentation through the B-type bosonization recalled below. On the other hand, the modes $W_{0,n}$ commute, they will be diagonal in the vertical presentation.

\para{Remark} It is equivalent, and sometimes more convenient, to present the algebra $\hat B_\infty$ using generators with integer indices $\COBp_{i,j}=\COB_{i+1/2,j-1/2}$. These generators obey the commutation relations
\begin{align}
\begin{split}
[\COBp_{i,j},\COBp_{k,l}]&=\d_{j+k}\COBp_{i,l}-\d_{i+l}\COBp_{k,j}\\
&+2C\left[\th(i+1/2)-\th(k+1/2)\right]\d_{j+k}\d_{i+l}\\
&+(-1)^{k+l}\left[\d_{i+k}\COBp_{l,j}-\d_{j+l}\COB_{i,k}\right]\\
&+(-1)^{k+l}2C\left[\th(l+1/2)-\th(i+1/2)\right]\d_{i+k}\d_{j+l},
\end{split}
\end{align}
together with the property $\COBp_{i,j}+(-1)^{i+j}\COB_{j,i}=-2C\d_{i,0}\d_{j,0}$. The algebra $B_\infty$ is recovered using the representation of level zero of $\glinf$ $\rho'_0(E_{r,s})=e_{r-1/2,-s-1/2}$ (it is a shift of the representation $\rho_0$ defined in the previous section). From \ref{def_COB}, we deduce that $\rho_0'(\COBp_{i,-j})=b_{i,j}$ with $b_{i,j}=e_{i,j}+(-1)^{i+j+1}e_{-j,-i}$.

\subsection{Representation on the Majorana fermion Fock space}
The algebra $\hat B_\infty$ can be represented on the Fock space of a self-adjoint fermion, also called neutral or Majorana fermion, with periodic (or Ramond) boundary conditions $\tphi(e^{2i\pi}z)=\tphi(z)$ \cite{Jimbo1983}. This field decomposes on integer modes \footnote{Instead, the Majorana fermion with anti-periodic (or Neuveu-Schwarz) boundary conditions, $\phi(e^{2i\pi}z)=-\phi(z)$, which is decomposed on half-integer modes, will appear in the representation of the subalgebra $\CW^D$ below (note, however, that the field will be multiplied by an extra power $z^{1/2}$ to get rid of the sign).} 
\begin{align}
\begin{split}\label{def_tphi}
&\tphi(z)=\sum_{k\in\mZ}z^{-k}\tphi_k,\quad \{\tphi_k,\tphi_l\}=(-1)^k\d_{k+l}\\
&\implies \{\tphi(z),\tphi(w)\}=\d(-z/w).
\end{split}
\end{align}
The modes $\tphi_k$ square to zero, except for the zero mode $(\tphi_0)^2=1/2$. The Fock space is obtained from the action of the modes $\tphi_{-k}$ for $k\geq0$ on the vacuum $\ketB{\vac}$ annihilated by the strictly positive modes, i.e. $\phi_k\ketB{\vac}=0$ for $k>0$. The normal ordering is usually defined by moving the positive modes to the right, but here we need to take into account the zero mode, and define instead 
\begin{align}\label{tphi_ordering}
\begin{split}
&:\tphi_k\tphi_l:=\tphi_k\tphi_l-\braB{\vac}\tphi_k\tphi_l\ketB{\vac},\\
&:\tphi(z)\tphi(w):=\tphi(z)\tphi(w)-\braB{\vac}\tphi(z)\tphi(w)\ketB{\vac},\\
\text{with}\quad&\braB{\vac}\tphi_k\tphi_l\ketB{\vac}=(-1)^{k}\d_{k+l}\th(k),\\
&\braB{\vac}\tphi(z)\tphi(w)\ketB{\vac}=\hf\dfrac{z-w}{z+w},
\end{split}
\end{align}
where the Heaviside function takes the value $\th(k)=1/2$ at $k=0$ (note that $:\tphi_k\tphi_l:=-:\tphi_l\tphi_k$).

The Majorana representation has level $1/2$, it is defined as
\begin{align}
\begin{split}\label{WB_phi}
&\rho^{(\tM)}(\COBp_{i,j})=(-1)^{j}:\tphi_i\tphi_{j}:-\hf\d_{i,0}\d_{j,0}\\
\implies&\rho^{(\tM)}(W^B_{m,n})=\sum_{k\in\mZ}(-1)^kq^{-nk}:\tphi_{m-k}\tphi_k:-\hf\d_{m,0}.
\end{split}
\end{align}
We denote this representation $\rho^{(\tM)}$ but omit the notation if no confusion ensues. It is sometimes useful to express the generators $W_{m,n}^B$ as a contour integral of the Majorana fermionic field,
\begin{equation}\label{WB_contour}
W^B_{m,n}=\oint\dfrac{dz}{2i\pi}z^{m-1}:\tphi(z)\tphi(-q^{n}z):-\hf\d_{m,0}.
\end{equation}
The modes act on the fermionic field as
\begin{align}
\begin{split}
&[W_{m,n}^B,\tphi(z)]=q^{-mn}z^m\tphi(q^{-n}z)+(-1)^{m+1}z^m\tphi(q^nz),
\end{split}
\end{align}
and, in particular, $[W^B_{m,0},\tphi(z)]=2\dodd{m}z^m\tphi(z)$ where $\dodd{m}$ is one if $m$ is odd and zero otherwise.

\para{Decomposition of the Dirac representation} As a subalgebra of $\CW$, the algebra $\CW^B$ also possesses a Dirac representation. From the action \ref{def_sB} of the automorphism $\s_B$ on the contour integral expression \ref{W_contour} of the generators $W_{m,n}$, we find for the subalgebra generators the formula
\begin{align}
\begin{split}
\rho^{(D)}(W_{m,n}^B)=&\oint{\dfrac{dz}{2i\pi}z^m\left(:\bpsi(z)\psi(q^nz):+q^{n}:\bpsi(-q^nz)\psi(-z):\right)}\\
&-\d_{m,0}.
\end{split}
\end{align}
It is well known that the charged Dirac fermion can be decomposed into two neutral Majorana fermions,
\begin{align}
\begin{split}
\psi(z)=\dfrac1{\sqrt{2}}\left(\tphi_1(-z)+i\tphi_2(-z)\right),\quad \bpsi(z)=\dfrac{z^{-1}}{\sqrt{2}}\left(\tphi_1(z)-i\tphi_2(z)\right).
\end{split}
\end{align}
Plugging this decomposition into the previous expression of the generators $W_{m,n}^B$, we find
\begin{align}
\begin{split}
\rho^{(D)}(W_{m,n}^B)=\oint\dfrac{dz}{2i\pi}z^{m-1}&\Big(:\tphi_1(z)\tphi_1(-q^nz):\\
&+:\tphi_2(z)\tphi_2(-q^nz):\Big)-\d_{m,0},
\end{split}
\end{align}
which corresponds to the fact that the Dirac representation of $\CW^B$ is not irreducible but decomposes into two Majorana representations as $\rho^{(D)}=\rho^{(\tM)}\otimes1+1\otimes\rho^{(\tM)}$.

\para{Bosonization} The Majorana fermion can be bosonized using a Heisenberg algebra $a_k$ with only odd modes $k$, and such that $[a_k,a_l]=2k\d_{k+l}$ (the extra factor $2$ in the r.h.s. is introduced for later convenience). In the mathematics literature, this is known as the twisted boson-fermion correspondence \cite{Jing1991}, but it is also sometimes simply referred as the bosonization of type B. The Majorana field is realized as
\begin{align}
\begin{split}\label{tphi_boson}
&\tphi(z)=\dfrac1{\sqrt{2}}\vdots e^{-\sum_{k\in\mZ}\dodd{k}\frac{z^{-k}}{k}a_k}\vdots\\
\implies &\tphi(z)\tphi(w)=\dfrac{z-w}{z+w}\vdots\tphi(z)\tphi(w)\vdots.,
\end{split}
\end{align}
where we denoted again $\vdots\cdots\vdots$ the bosonic normal ordering. Note that the adjoint action $a_k^\dagger=a_{-k}$ corresponds to $\tphi(z)^\dagger=\tphi(-z^{-1})$ in the case of Majorana, and $\psi(z)^\dagger=z^{-1}\bpsi(z^{-1})$ in the case of Dirac. Computing $W_{k,0}^B$ in the Majorana representation \ref{WB_phi} leads to identify the Heisenberg algebras $a_k=W_{k,0}^B$. The expression for the other generators is found by performing the bosonization on the contour integral formula \ref{WB_contour},
\begin{equation}
W_{m,n}^B=\hf\dfrac{1+q^n}{1-q^n}\oint\dfrac{dz}{2i\pi}z^{m-1}\vdots e^{-\sum_{k\in\mZ}\dodd{k}\frac{z^{-k}}{k}(1-q^{-nk})a_k}\vdots-\dfrac{\d_{m,0}}{1-q^n}.
\end{equation} 
In this way, we recover the vertex representation for the algebra $\CW^B$ proposed in \cite{Lebedev1992}.

\subsection{Vertical presentation}
The Fock space for the periodic Majorana fermion is spanned by the states $\ketB{\l}$ labeled by the symmetric partitions $\l=\l'$. These states are defined using the hook decomposition of the Young diagram, assigning the action of the modes $\tphi_{-k_i}$ to the $i$th hook of length $h_i=2k_i+1$,
\begin{equation}
\ketB{\l}=\tphi_{-k_1}\cdots\tphi_{-k_{d(\l)}}\ketB{\vac},\quad k_i\in\mZ.
\end{equation}
The hooks are strictly ordered, and thus $k_1>k_2>\cdots>k_{d(\l)}$.\footnote{Alternatively, following a well-known bijection, the set of hooks can be seen as defining a strict partition.} In this setting, the action of the zero mode $\tphi_0$ is to add/remove a box on the diagonal (with a multiplication by a factor $1/2$ if the box is removed). Dual states are obtained from the action of the adjoint modes $\tphi^\dagger_{-k}=(-1)^k\tphi_k$ (with $k>0$) on the dual vacuum $\braB{\vac}$, they are orthonormal: $\scalarB{\l}{\m}=\d_{\l,\mu}$.

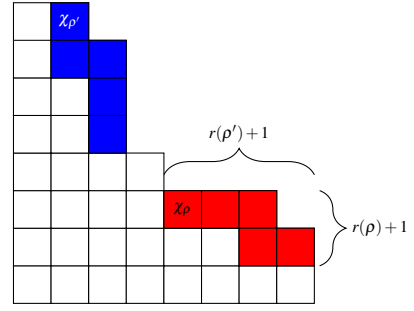
\begin{figure}
\begin{center}
\begin{tikzpicture}[scale=0.5]
\draw [fill=white] (1,1) rectangle (2,2);
\draw [fill=white] (1,2) rectangle (2,3);
\draw [fill=white] (2,1) rectangle (3,2);
\draw [fill=white] (1,3) rectangle (2,4);
\draw [fill=white] (3,1) rectangle (4,2);
\draw [fill=white] (1,4) rectangle (2,5);
\draw [fill=white] (4,1) rectangle (5,2);
\draw [fill=white] (1,5) rectangle (2,6);
\draw [fill=white] (5,1) rectangle (6,2);
\draw [fill=white] (1,6) rectangle (2,7);
\draw [fill=white] (6,1) rectangle (7,2);
\draw [fill=white] (1,7) rectangle (2,8);
\draw [fill=white] (7,1) rectangle (8,2);
\draw [fill=white] (1,8) rectangle (2,9);
\draw [fill=white] (8,1) rectangle (9,2);
\draw [fill=white] (2,2) rectangle (3,3);
\draw [fill=white] (2,3) rectangle (3,4);
\draw [fill=white] (3,2) rectangle (4,3);
\draw [fill=white] (2,4) rectangle (3,5);
\draw [fill=white] (4,2) rectangle (5,3);
\draw [fill=white] (2,5) rectangle (3,6);
\draw [fill=white] (5,2) rectangle (6,3);
\draw [fill=white] (2,6) rectangle (3,7);
\draw [fill=white] (6,2) rectangle (7,3);
\draw [fill=white] (2,7) rectangle (3,8);
\draw [fill=white] (7,2) rectangle (8,3);
\draw [fill=white] (2,8) rectangle (3,9);
\draw [fill=white] (8,2) rectangle (9,3);
\draw [fill=white] (3,3) rectangle (4,4);
\draw [fill=white] (3,4) rectangle (4,5);
\draw [fill=white] (4,3) rectangle (5,4);
\draw [fill=white] (3,5) rectangle (4,6);
\draw [fill=white] (5,3) rectangle (6,4);
\draw [fill=white] (3,6) rectangle (4,7);
\draw [fill=white] (6,3) rectangle (7,4);
\draw [fill=white] (3,7) rectangle (4,8);
\draw [fill=white] (7,3) rectangle (8,4);
\draw [fill=white] (4,4) rectangle (5,5);
\draw [fill=red] (5,3) rectangle (6,4);
\draw [fill=red] (6,3) rectangle (7,4);
\draw [fill=red] (7,3) rectangle (8,4);
\draw [fill=red] (7,2) rectangle (8,3);
\draw [fill=red] (8,2) rectangle (9,3);
\node [scale=.7] at (5.5,3.5) {$\chi_{\rho}$};
\draw [fill=blue] (3,5) rectangle (4,6);
\draw [fill=blue] (3,6) rectangle (4,7);
\draw [fill=blue] (3,7) rectangle (4,8);
\draw [fill=blue] (2,7) rectangle (3,8);
\draw [fill=blue] (2,8) rectangle (3,9);
\node [white,scale=.7] at (2.5,8.5) {$\chi_{\rho'}$};
\draw [decorate,decoration={brace,amplitude=10pt,mirror,raise=4pt},xshift=-4pt,yshift=0pt] (9,2) -- (9,4) node [black,midway,right,xshift=14pt,scale=.7]  {$r(\rho)+1$};
\draw [decorate,decoration={brace,amplitude=10pt,mirror,raise=4pt},xshift=0pt,yshift=4pt] (9,4) -- (5,4) node [black,midway,above,yshift=14pt,scale=.7]  {$r(\rho')+1$};
\end{tikzpicture}
\end{center}
\caption{Example of a symmetric Young diagram with a strip $\rho\in R_5(\l)$ (red), and its transposed $\rho'$ (blue), with $\chi_{\rho}=q^2$ and $\chi_{\rho'}=q^{-6}$.}
\label{fig3}
\end{figure}

In order to write down the vertical presentation, we need to introduce again several notations. To keep the Young diagram symmetric, the horizontal strips will be added/removed in pairs. For this purpose, we define the sets $A_m^B(\l)$ (resp. $R_m^B(\l)$) of strips with $m$-boxes lying \textbf{strictly below the diagonal} that can be added to (resp. removed from) the symmetric partition $\l$. Such strips $\rho$ have a unique transposed $\rho'\in A_m(\l)$ (resp. $\rho'\in R_m(\l)$) lying strictly above the diagonal. As before, the content $\chi_\rho$ of a strip $\rho\in A_m^B(\l)\cup R_m^B(\l)$ is defined as the content of its box in the top left corner. Thus, the transposed strip has the content $\chi_{\rho'}=q^{1-m}\chi_\rho^{-1}$. The modes $W_{m,n}^B$ act by adding or removing the pairs $(\rho,\rho')$ and we denote the corresponding state $\ketB{\l\pm \rho}$. We keep the notation $r(\rho)$ for the number of rows that the strip $\rho$ occupies minus one (see the example of figure \ref{fig3}).

In addition, the vertical action also involves the addition/removal of a pair of hooks. We denote $A_m^{HP}(\l)$ (resp. $R_m^{HP}(\l)$) the set of hook pairs with a total of $2m$ boxes, that can be added to (resp. removed from) the symmetric diagram $\l$. These hooks are not necessarily adjacent, and they can be inserted/removed anywhere in the diagram. For $\rho\in A_m^{HP}(\l)\cup R_m^{HP}(\l)$, we denote $\ketB{\l\pm \rho}$ the state corresponding to the addition/removal of the hook pair $\rho$. For a hook pair $\rho$, let $k(\rho)$ be the integer such that the smaller hook has $2k(\rho)+1$ boxes. The content is defined as the box content of the box in the top left corner of the longest hook, and thus $\chi_\rho=q^{k(\rho)-m+1}$. Finally, we denote $r(\rho)+1$ the number of hooks of the diagram $\l$ lying between the two hooks to be inserted/removed (we omit the dependence in the partition $\l$ for this notation).

The action of the operators $W^B_{m,n}$ on the states $\ketB{\l}$ is obtained by a direct, but tedious, calculation,
\begin{align}\label{WB_vert}
\begin{split}
&W^B_{0,n}\ketB{\l}=\left((1-q^n)\sum_{\sAbox\in\l_-}(q^{-n}\chi_{\sAbox}^n+\chi_{\sAbox}^{-n})-\hf \right)\ketB{\l},\\
&W^B_{m,n}\ketB{\l}=\\
&\sum_{\rho\in R_m^B(\l)}\left(q^{-(m-1)n}\chi_\rho^{-n}+(-1)^{m+1}q^{-n}\chi_\rho^n)\right)(-1)^{r(\rho)}\ketB{\l-\rho}\\
&+\sum_{\rho\in R_{m+1}^{HP}(\l)}\left(1-\hf\d_{k(\rho),0}\right)\left(q^{-mn}\chi_\rho^{-n}+(-1)^{m+1}\chi_\rho^{n}\right)\\
&\qquad\qquad(-1)^{r(\rho)+k(\rho)+m}\ketB{\l-\rho},\\
&W^B_{-m,n}\ketB{\l}=\\
&\sum_{\rho\in A_{m+1}^{HP}(\l)}\left(q^{mn}\chi_\rho^n+(-1)^{m+1}\chi_\rho^{-n}\right)(-1)^{r(\rho)+k(\rho)+1}\ketB{\l+\rho}\\
&+\sum_{\rho\in A_m^B(\l)}\left(1-\hf\d_{\chi_\rho,q}\right)\left(q^n\chi_\rho^{-n}+(-1)^{m+1}q^{(m-1)n}\chi_\rho^n\right)\\
&\qquad\qquad(-1)^{r(\rho)}\ketB{\l+\rho}.
\end{split}
\end{align}
In the first line, the sum is performed on the set of boxes $\l_-$ lying strictly below the diagonal. The Kronecker deltas provide an extra factor $1/2$ when $\tphi_0$ is involved. These rules might appear complicated, but they can be easily implemented on a computer program. In particular, we have for $a_k=W_{k,0}^B$ with $k$ odd:
\begin{align}\label{ak_B}
\begin{split}
a_k\ketB{\l}&=2\!\!\!\sum_{\rho\in R_k^B(\l)}(-1)^{r(\rho)}\ketB{\l-\rho}\\
&+2\!\!\!\sum_{\rho\in R_{k+1}^{HP}(\l)}\left(1-\hf\d_{k(\rho),0}\right)(-1)^{r(\rho)+k(\rho)+k}\ketB{\l-\rho},\\
a_{-k}\ketB{\l}&=2\!\!\!\sum_{\rho\in A_k^B(\l)}\left(1-\hf\d_{\chi_\rho,q}\right)(-1)^{r(\rho)}\ketB{\l+\rho}\\
&+2\!\!\!\sum_{\rho\in A_{k+1}^{HP}(\l)}(-1)^{r(\rho)+k(\rho)+1}\ketB{\l+\rho}.
\end{split}
\end{align}

It might be surprising to notice that the action of the operators $W_{m,n}^B$ produces states with a different number of boxes: the addition/removal of two strips of length $m$ changes the number of boxes by $2m$ while the addition/removal of a pair of hooks modifies it by $2m+2$. However, in both cases, the number of boxes strictly below the diagonal $|\l_-|$ varies by $\pm m$. This property can be formally encoded in the action of a grading operator $d_B$ such that $[d_B,W_{m,n}^B]=-mW_{m,n}^B$,
\begin{align}
\begin{split}
&d_B=\hf\sum_{k\in\mZ}(-1)^kk:\tphi_{-k}\tphi_k:\implies [d_B,\tphi_k]=-k\tphi_k,\\
&d_B\ketB{\l}=|\l_-|\ketB{\l},\quad \a^{d_B}W_{m,n}^B\a^{-d_B}=\a^{-m}.
\end{split}
\end{align} 
Furthermore, the number of hooks varies by $0,\pm2$ under the action of the generators $W_{m,n}^B$. Thus, the Majorana representation is not irreducible, but it decomposes into an action on symmetric partitions with either an even or an odd number of hooks.\footnote{Since $\l$ is symmetric, $|\l|=2|\l_-|+d(\l)$ and the parity of the number of hooks is the same as the parity of the number of boxes.} This is the reason for introducing two sets of symmetric polynomials $b_\l(x)$ and $b_\l^\ast(x)$ below.

\subsection{Symmetric polynomials}
The bosonic formula \ref{Schur_corr} for the Schur polynomials can be generalized to the Majorana fermion and its bosonization, it defines two sets of polynomials labeled by symmetric partitions,
\begin{align}
\begin{split}\label{def_bl}
&b_\l(x)=\braB{\vac}e^{\sum_{k>0}\dodd{k}\frac{p_k(x)}{k}a_k}\ketB{\l},\\
&b_\l^\ast(x)=\braB{\Abox}e^{\sum_{k>0}\dodd{k}\frac{p_k(x)}{k}a_k}\ketB{\l}.
\end{split}
\end{align}
The action of the modes $a_k$ on the states $\ketB{\l}$ is given by the vertical formula \ref{ak_B}, it produces only states labeled by partitions with a strictly smaller number of boxes, and so the expansion of the r.h.s. is finite. We provide here a few examples:
\begin{align}
\begin{split}
&b_\vac=1,\quad b_{2^2}=p_1,\quad b_{321}=p_1^2,\quad b_{3^22}=\dfrac2{3}p_1^3-\dfrac23p_3,\\
&b_{421^2}=\dfrac2{3}p_1^3+\dfrac13p_3,\quad b_{4321}=\dfrac2{3}p_1^4-\dfrac23p_1p_3,\\
&b_{1}^\ast=\hf,\quad b_{21}^\ast=p_1,\quad b_{31^2}^\ast=p_1^2,\quad b_{3^3}^\ast=\dfrac13p_1^3-\dfrac13p_3,\\
&b_{41^3}^\ast=\dfrac23p_1^3+\dfrac13p_3,\quad b_{51^4}=\dfrac13p_1^4+\dfrac23p_1p_4.
\end{split}
\end{align}
Let us stress that these polynomials do not depend on the parameter $q$, and that they involve only odd power sums by definition. Inserting the grading operator in the correlator, one can show, like in the Schur case, that they are homogeneous polynomials of degree $|\l_-|$. Moreover, it is easily seen from the action of $a_k$ that $b_\l(x)$ is non-zero for $|\l|$ even and $b_\l^\ast(x)$ for $|\l|$ odd. In fact, they can be seen as special cases of the more general skew polynomials
\begin{equation}
b_{\l/\mu}(x)=\braB{\mu}e^{\sum_{k>0}\dodd{k}\frac{p_k(x)}{k}a_k}\ketB{\l}.
\end{equation} 
From the action \ref{ak_B} of $a_k$, we observe that this expression is non-vanishing only if the Young diagram $\mu$ is contained in $\l$.

In the following, we focus mostly on the polynomial $b_\l(x)$. This polynomials is, up to normalization, a Q-Schur function  $Q_{s(\l)}=2^{\sharp}b_\l$ for the strict partition $s(\l)$ uniquely associated to the symmetric partition $\l$ and defined by the hook lengths, $s(\l)=((h_1-1)/2,(h_2-1)/2,\cdots)$.\footnote{We have $\sharp=\lfloor\ell(s(\l))/2\rfloor$.} This identification follows from the fermionic expression,
\begin{align}
\begin{split}\label{bl_fields}
&b_\l^{(N)}(x)=2^{N/2}\braB{\vac}\vdots\tphi(-x_1^{-1})\tphi(-x_2^{-1})\cdots\tphi(-x_N^{-1})\vdots\ketB{\l}\\
&=2^{N/2}\D_B(x)^{-1}\braB{\vac}\tphi(-x_1^{-1})\tphi(-x_2^{-1})\cdots\tphi(-x_N^{-1})\ketB{\l}\\
&\text{with}\quad \D_B(x)=2^{N/2}\braB{\vac}\tphi(-x_1^{-1})\tphi(-x_2^{-1})\cdots\tphi(-x_N^{-1})\ketB{\vac},\\
&\qquad\qquad\quad\ =\prod_{\superp{i,j=1}{i<j}}^N\dfrac{x_j-x_i}{x_j+x_i}
\end{split}
\end{align}
where the polynomials $b_\l(x)$ with infinitely many variables are obtained as an inductive limit of the polynomials $b_\l^{(N)}(x)$ with $N$ variables. This fermionic expression is the equivalent of the formula $s_\l(x)=\det(x_i^{\l_j+N-j})/\D(x)$ for the Schur polynomials, it allows us to rewrite the polynomials in terms of Pfaffians \cite{You1989,Nimmo1990,Jing1991}.\footnote{It is also worth mentioning the decomposition of Schur polynomials into a sum of products of Q-functions derived in \cite{Harnad2008,Harnad2020} using the relation between Dirac and Majorana fermions.} Note, however, that not all Q-Schur functions $Q_\l(x)$ are obtained in this way since we have the restriction that $\l$ is a strict partition (otherwise, the decomposition of $Q_\l(x)$ may also contain even power sums). We will attempt to generalize these fermionic formulas to the case of $\hat C_\infty$ and $\hat D_\infty$ in the next sections.

Just like in the case of Schur polynomials, the introduction of the modes $W_{k,0}^B$ inside the correlators defining $b_\l(x)$ and $b_\l^\ast(x)$ leads to the derivation of Pieri-like rules
\begin{align}
\begin{split}
&p_k(x)b_\l(x)=\sum_{\rho\in A_k^B(\l)}\left(1-\hf\d_{\chi_\rho,q}\right)(-1)^{r(\rho)}b_{\l+\rho+\rho'}(x)\\
&+\sum_{\rho\in A_{k+1}^{HP}(\l)}(-1)^{r(\rho)+k(\rho)+1}b_{\l+\rho}(x),\\
&k\dfrac{\p}{\p p_k(x)}b_\l(x)=\ 2\!\!\!\sum_{\rho\in R_k^B(\l)}(-1)^{r(\rho)}b_{\l-\rho-\rho'}(x)\\
&+2\!\!\!\sum_{\rho\in R_{k+1}^{HP}(\l)}\left(1-\hf\d_{k(\rho),0}\right)(-1)^{r(\rho)+k(\rho)+k}b_{\l-\rho}(x),
\end{split}
\end{align}
and the same rules for $b_\l^\ast(x)$. The rules obtained here differ from the usual Pieri rules for Q-Schur polynomials for which the multiplication by the complete symmetric functions $h_k$ is considered instead. This explains why the r.h.s. is more complicated, involving also the addition/removal of pairs of hooks.

In addition, the introduction of the dual modes $W_{0,n}^B$ also leads to a q-difference equation. Since the derivation is a bit lengthy, we kept it in appendix \ref{AppB}. We found
\begin{align}
\begin{split}\label{B_qdiff}
&\sum_i\left(B_i^{(n)}(x)T_{q^{-n},x_i}-B_i^{(-n)}(x)T_{q^n,x_i}\right)\ b_\l(x)\\
=&(1-q^n)\sum_{\sAbox\in\l_-}(q^{-n}\chi_{\sAbox}^n+\chi_{\sAbox}^{-n})\ b_\l(x),
\end{split}
\end{align}
with
\begin{equation}\label{def_Ai}
B_i^{(n)}(x)=\prod_{j\neq i}\dfrac{(x_i+x_j)(x_i-q^nx_j)}{(x_i-x_j)(x_i+q^nx_j)}.
\end{equation} 
The polynomials $b_\l^\ast(x)$ satisfy the same q-difference equation. Let us stress that not all Q-Schur functions satisfy this q-difference equation, but only those labeled by a strict partition. To the knowledge of the author, this q-difference equation is new.\footnote{In particular, it appears to be unrelated to the Koornwinder operators associated to the BC$_n$ root system \cite{Koornwinder1992}, and defining the BC$_n$ Ruijsenaars model \cite{vanDiejen1995}, that depends on the six parameters $a,b,c,d,q,t$,
\begin{align}
\begin{split}
\CD=&\sum_id(x_i)\prod_{j\neq i}\dfrac{(1-tx_ix_j)(1-tx_i/x_j)}{(1-x_ix_j)(1-x_i/x_j)}(T_{q,x_i}-1)\\
&+\sum_id(x_i^{-1})\prod_{j\neq i}\dfrac{(1-t/(x_ix_j))(1-tx_j/x_i)}{(1-1/(x_ix_j))(1-x_j/x_i)}(T_{q^{-1},x_i}-1)\\
\text{with}\quad & d(z)=\dfrac{(1-az)(1-bz)(1-cz)(1-dz)}{(abcd)^{1/2}q^{-1/2} t^{n-1}(1-x_i^2)(1-qx_i^2)}
\end{split}
\end{align}
Indeed, the difference operator \ref{B_qdiff} involves the linearized versions $x_i-x_j$, $x_i+x_j$ of the ratios $x_i/x_j$ and products $x_ix_j$ appearing in the Koornwinder operator, while keeping the same difference operators $T_{q^{\pm1},x_i}$.}


\para{Hamiltonian} Like the Schur polynomials, the polynomials $b_\l(x)$ satisfy a q-difference equation and are independent of $q$. Expanding at $q=1$, it is possible to derive a tower of differential equations satisfied by these polynomials. However, since the q-difference equation is antisymmetric under the exchange $q\to q^{-1}$, only odd powers of $\a$ remain in the expansion of $q^n=e^{-\a}$. The expansion easily follows from the following fact,
\begin{align}
\begin{split}
\CT^B=&\sum_i\left(B_i^{(n)}(x)T_{q^{-n},x_i}-B_i^{(-n)}(x)T_{q^n,x_i}\right)\\
=&\D_B(x)^{-1}\sum_i\left(T_{q^{-n},x_i}-T_{q^n,x_i}\right)\D_B(x),
\end{split}
\end{align}
with $\D_B(x)$ defined in \ref{bl_fields}, and we find $\CT_B=2\a\sum_iD_i+\a^3\CH_3+O(\a^5)$. The first order term corresponds to the homogeneity property of the polynomials $b_\l(x)$. Since the term of order $O(\a^2)$ is missing, we do not have a Hamiltonian in the usual sense, i.e. with a kinetic term involving second order differentials.\footnote{Note the surprising fact that if we take instead the symmetric combination
\begin{equation}
\sum_i\left(B_i^{(n)}(x)T_{q^{-n},x_i}+B_i^{(-n)}(x)T_{q^n,x_i}\right)=2N+\a^2\CH_2+O(\a^4), 
\end{equation} 
we find at the subleading order the Hamiltonian of the Calogero-Sutherland model with periodic boundary conditions \cite{Wang1996} (plus some terms independent of derivatives),
\begin{align}
\begin{split}
\CH_2=&\sum_iD_i^2+\sum_{\superp{i,j}{i<j}}\dfrac{x_i+x_j}{x_i-x_j}(D_i-D_j)-\sum_{\superp{i,j}{i<j}}\dfrac{x_i-x_j}{x_i+x_j}(D_i-D_j)\\
&+\sum_i\D_B(x)^{-1}D_i^2(\D_B(x)).
\end{split}
\end{align}\protect\trick.} The first non-trivial equation contains derivatives of order three,
\begin{align}
\begin{split}
\CH_3&=\dfrac13\sum_i D_i^3+\hf\sum_{i<j}\dfrac{x_i+x_j}{x_i-x_j}(D_i^2-D_j^2)-\hf\sum_{i<j}\dfrac{x_i-x_j}{x_i+x_j}(D_i^2-D_j^2)\\
&-2\sum_{i<j}\dfrac{x_ix_j}{(x_i+x_j)^2}(D_i+D_j)+4\sum_{\superp{i,j,k}{i\neq j\neq k}}\dfrac{x_i^2x_jx_k}{(x_i^2-x_j^2)(x_i^2-x_k^2)}D_i\\
&+\dfrac13\sum_i \D_B(x)^{-1}D_i^3(\D_B(x)).
\end{split}
\end{align}

\section{$C$-type orbifold and $\spinf$}
The algebra $\hat C_\infty$, also called $\spinf$, is a central extension of the algebra of infinite matrices with integer indices $c_{i,j}$ that obey the relation $c_{i,j}=(-1)^{i+j+1}c_{1-i,1-j}$. It can be defined using the generators $\COC_{r,s}$ with half-integer indices, and the central element $C$, satisfying the relations
\begin{align}\label{def_spinf}
\begin{split}
&[\COC_{r,s},\COC_{t,u}]=\d_{s+t}\COC_{r,u}-\d_{r+u}\COC_{t,s}+2C(\th(r)-\th(t))\d_{s+t}\d_{r+u}\\
&+(-1)^{t+u}\left[\d_{r+t}\COC_{u,s}-\d_{s+u}\COC_{r,t}+2C(\th(u)-\th(r))\d_{s+u}\d_{r+t}\right].
\end{split}
\end{align}
This algebra is a subalgebra of $\glinf$ generated by the linear combinations
\begin{equation}\label{def_COC}
\COC_{r,s}=E_{r,s}+(-1)^{r+s+1}E_{s,r}.
\end{equation}
The algebra $C_\infty$ is obtained using the representation $\rho_0$ of level $0$ of $\mathfrak{gl}(\infty)$ defined in the section \ref{sec_def_rho} as $\rho_0(\COC_{i-1/2,-j+1/2})=c_{i,j}$ with $c_{i,j}=e_{i,j}+(-1)^{i+j+1}e_{1-j,1-i}$.

The transformation \ref{rel_W_E} applied to the linear combinations \ref{def_COC} defines the generators
\begin{equation}\label{def_tWV}
W^C_{m,n}=\sum_{r\in\mZ+1/2}q^{-(r+1/2)n}\COC_{m-r,r}
\end{equation} 
of a subalgebra of the quantum $\Winf$ algebra that we denote $\CW^C$. This subalgebra is again a $\mZ_2$-orbifold, it is obtained from the following involutive automorphism \cite{Lebedev1992}
\begin{equation}\label{def_sC}
\s_C:\ W_{m,n}\to (-1)^{m+1}q^{-(m+1)n}W_{m,-n},\quad C\to C.
\end{equation} 
The subalgebra $\CW^C$ is generated by the central element $(1+\s_C)C=2C$ and the projections $W^C_{m,n}=(1+\s_C)W_{m,n}$ that coincide with the definitions \ref{def_tWV}. They satisfy the commutation relations
\begin{align}
\begin{split}\label{com_WC}
[W_{m,n}^C,W_{m',n'}^C]&=(q^{m'n}-q^{mn'})\left(W^C_{m+m',n+n'}+2C\dfrac{\d_{m+m'}}{1-q^{n+n'}}\right)\\
&+(-1)^{m'+1}q^{-(m'+1)n'}(q^{m'n}-q^{-mn'})W^C_{m+m',n-n'}\\
&+(-1)^{m'+1}q^{-(m'+1)n'}\dfrac{q^{m'n}-q^{-mn'}}{1-q^{n-n'}}2C\d_{m+m'}.
\end{split}
\end{align}
As in the case of $\hat B_\infty$, the modes $W_{m,0}$ with $m$ even are projected out. The modes $W_{m,0}^C$ with $m$ odd form an Heisenberg subalgebra, $[W_{m,0}^C,W_{m',0}^C]=4Cm\d_{m+m'}$, while the modes $W_{0,n}^C$ commute. They will be used in the horizontal/vertical presentations respectively.

\subsection{Representation on the symplectic boson Fock space}
The algebra $\spinf$ is known to possess a representation of level $-1/2$ on the Fock space of a symplectic boson with anti-periodic boundary conditions,
\begin{equation}
\hphi(z)=\sum_{r\in\mZ+1/2}z^{-r-1/2}\hphi_r,\quad [\hphi_r,\hphi_s]=(-1)^{r-1/2}\d_{r+s}.
\end{equation}
The Fock space is built from the action of negative modes on the vacuum $\ketC{\vac}$ annihilated by positive modes.  The normal ordering is defined by moving to the right the positive modes, so that\footnote{These algebraic relations expressed in terms of the bosonic field $\hphi(z)$ read
\begin{equation}
[\hphi(z),\hphi(w)]=z^{-1}\d(-z/w),\quad \bra{\vac}\hphi(z)\hphi(w)\ket{\vac}=\dfrac1{z+w}.
\end{equation}\protect\trick.}
\begin{equation}
:\hphi_r\phi_s:=\hphi_r\hphi_s-\bra{\vac}\hphi_r\hphi_s\ket{\vac},\quad \bra{\vac}\hphi_r\hphi_s\ket{\vac}=(-1)^{r-1/2}\th(r)\d_{r+s}.
\end{equation} 
We denote the representation on this Fock space as $\rho^{(SB)}$, but omit the notation if no confusion ensues. The generators $\COC_{r,s}$ and $W_{m,n}^C$ of respectively $\hat C_\infty$ and $\CW^C$ are represented by
\begin{align}
\begin{split}\label{WC_phi}
&\rho^{(SB)}(\COC_{r,s})=(-1)^{s-1/2}:\hphi_r\hphi_s:,\\
&\rho^{(SB)}(W^C_{m,n})=\sum_{r\in\mZ+1/2}(-1)^{s-1/2}q^{-(r+1/2)n}:\hphi_{m-r}\hphi_r:.
\end{split}
\end{align}
Alternatively, we can write the generators of $\CW^C$ in the form of the contour integrals
\begin{equation}
\rho^{(SB)}(W^C_{m,n})=-\oint\dfrac{dz}{2i\pi}z^m:\hphi(z)\hphi(-q^nz):,
\end{equation}
they act on the bosonic field as follows,
\begin{align}
\begin{split}
&[W^C_{m,n},\hphi(z)]=(-1)^{m+1}z^m\hphi(q^nz)+q^{-n(m+1)}z^m\hphi(q^{-n}z),
\end{split}
\end{align}
in particular $[W_{m,0}^C,\hphi(z)]=2\dodd{m}z^m\hphi(z)$.

\para{Decomposition of the $\b\g$-representation} Since $\CW^C$ is a subalgebra of $\CW$, it can also be represented on the Fock space of the bosonic ghosts $(\beta,\gamma)$. The expression of the generators $W_{m,n}^C$ follows from the action of the involution \ref{def_sC},
\begin{equation}
\rho^{(\b\g)}(W^C_{m,n})=\oint\dfrac{dz}{2i\pi}z^m\left(:\b(z)\g(q^nz):+:\b(-q^nz)\g(-z):\right).
\end{equation} 
It is well-known that the $\b\g$-system is equivalent to two symplectic bosons under the linear relations
\begin{equation}
\b(z)=\dfrac1{\sqrt{2}}\left(\hphi_1(-z)+\hphi_2(z)\right),\quad \g(z)=\dfrac1{\sqrt{2}}\left(\hphi_1(z)-\hphi_2(-z)\right),
\end{equation}
that allow us to rewrite the representation of the generators for $\CW^C$ in the form
\begin{align}
\begin{split}
&\rho^{(\b\g)}(W^C_{m,n})=\\
&-\oint\dfrac{dz}{2i\pi}z^m\left((-1)^{m}:\hphi_1(z)\hphi_1(-q^nz):+:\hphi_2(z)\hphi_2(-q^nz):\right).
\end{split}
\end{align}
This formula simply expresses the fact that the $(\b\g)$-representation can be decomposed into two symplectic boson representations as $\rho^{(\b\g)}=\trho^{(SB)}\otimes1+1\otimes\rho^{(SB)}$ where $\trho^{(SB)}$ denotes another representation of level $-1/2$ on a symplectic boson Fock space, differing from $\rho^{(SB)}$ by a choice of signs, namely $\trho^{(SB)}(\COC_{r,s})=(-1)^{r+1/2}:\hphi_r\hphi_s:$, which is equivalent to the multiplication of the generators $W_{m,n}^C$ by a factor $(-1)^m$.

\para{Twisted bosonization} We review here the bosonization performed by Van de Leur, Orlov and Shiota in \cite{VanDeLeur2012} (following a suggestion from Date, Jimbo, Kashiwara and Miwa \cite{Date1981}). It is sometimes called the \textit{twisted bosonization}, and describes, in fact, a correspondence between the symplectic boson and a boson+fermion system.\footnote{It differs from the \textit{untwisted bosonization} introduced by Anguelova in \cite{Anguelova2017a,Anguelova2017}.} The bosonic modes correspond to the modes $W_{m,n}^C$ with $n=0$ and $m$ odd that form the Heisenberg subalgebra. In this section, we denote $a_k=W_{k,0}^C$ for short, we have $[a_k,a_l]=-2k\d_{k+l}$ for this representation of level $C=-1/2$. Following \cite{VanDeLeur2012}, we introduce a dressed version of the bosonic field,
\begin{equation}\label{def_th}
\chi(z)=V_-(z)\hphi(z)V_+(z),\quad V_\pm(z)=e^{\mp\sum_{k>0}\dodd{k}\frac{z^{\mp k}}{k}a_{\pm k}}.
\end{equation} 
The dressing is chosen such that the field $\chi(z)$ commutes with the modes $a_k$. From the commutator $[a_k,\hphi(z)]=2z^k\hphi(z)$, we deduce the exchange relations
\begin{align}
\begin{split}
&V_\pm(z)\hphi(w)=\pm\left(\dfrac{z-w}{z+w}\right)^{\pm1}\hphi(w)V_\pm(z),\\
&V_+(z)V_-(w)=\dfrac{z+w}{z-w}V_-(w)V_+(z),
\end{split}
\end{align}
which leads to
\begin{equation}
\chi(z)\chi(w)=\dfrac{z-w}{z+w}V_-(z)V_-(w)\hphi(z)\hphi(w)V_+(z)V_+(w).
\end{equation} 
After a short computation that we chose not to reproduce here as it is found in \cite{VanDeLeur2012}, it is possible to show that the field $\chi(z)$ anticommutes with itself, and thus should be thought of as a fermionic field. Decomposing the field on half-integer modes, the anticommutator reads\footnote{The expression of the commutator of the field is more involved, it read $\{\chi(z),\chi(w)\}=z^{-1}\D(z/w)$, where we introduced the distribution $\D(z)=\d(-z)-2\d'(-z)$ with
\begin{equation}
\d'(z)=\sum_{k\in\mZ}kz^k=z\p_z\d(z)\implies f(z)\d'(z/a)=f(a)\d'(z/a)-af'(a)\d(z/a).
\end{equation}
In fact, the distribution $\D(z)$ can also be obtained from the difference of the expansions in powers of $z^{\pm1}$ of the vacuum expectation values $\braC{\vac}\chi(z)\chi(w)\ketC{\vac}$,
\begin{equation}
\left[\dfrac{z-w}{(z+w)^2}\right]_+-\left[\dfrac{z-w}{(z+w)^2}\right]_-=z^{-1}\D(z/w).
\end{equation}\protect\trick.}
\begin{equation}
\chi(z)=\sum_{r\in\mZ+1/2}z^{-r-1/2}\chi_r,\quad \{\chi_r,\chi_s\}=2r(-1)^{r-1/2}\d_{r+s}.
\end{equation} 
The symplectic boson vacuum state is annihilated by both bosonic and fermionic positive modes, $a_k\ketC{\vac}=\chi_r\ketC{\vac}=0$ for $k,r>0$. For the bosonic modes, it follows from the vertical presentation given below, while for the fermionic modes, it is seen using the definition \ref{def_th} after expanding in powers of $z$. It leads to define a fermionic normal-ordering by moving positive modes to the right. Then
\begin{align}
\begin{split}
&:\chi(z)\chi(w):=\chi(z)\chi(w)-\braC{\vac}\chi(z)\chi(w)\ketC{\vac},\\
&\braC{\vac}\chi(z)\chi(w)\ketC{\vac}=\dfrac{z-w}{(z+w)^2}.
\end{split}
\end{align}

Thus, the horizontal presentation of the representation involves not only the bosonic modes $a_k$, but also the fermionic ones $\chi_r$. As we shall see, this fact renders the analysis of the corresponding symmetric polynomials much more complicated. The horizontal presentation can be written in terms of the currents
\begin{equation}\label{def_wnC}
w_n^C(z)=(1-q^n)\sum_{m\in\mZ}z^{-m}W_{m,n}^C+2C.
\end{equation}
In the symplectic boson representation, the currents reads $w_n^C(z)=-(1-q^n)z\hphi(z)\hphi(-q^nz)$. After the twisted bosonization, they become
\begin{align}
\begin{split}\label{wnC_twisted}
w_n^C(z)=-&\dfrac{(1-q^n)^2}{1+q^n}ze^{-\sum_{k>0}\dodd{k}(1-q^{nk})\frac{z^k}{k}a_{-k}}\\
&\chi(z)\chi(-q^nz)e^{\sum_{k>0}\dodd{k}(1-q^{-nk})\frac{z^{-k}}{k}a_{k}}.
\end{split}
\end{align}

\subsection{Vertical presentation}
There are several ways to label the PBW basis of the symplectic boson Fock space. Here, we follow again the reference \cite{VanDeLeur2012} and use the set of partitions with odd parts, denoted OP, and represented as Young diagrams with an odd number of boxes in each column. The basis consists of the vectors
\begin{equation}
\ketC{\l}=\hphi_{-\l_1/2}\cdots \hphi_{-\l_\ell/2}\ketC{\vac},\quad \l\in\text{OP},
\end{equation} 
and the dual basis $\braC{\l}$ is obtained from the action of the adjoint modes $\hphi^\dagger_{-r}=(-1)^{r+1/2}\hphi_r$ with $r>0$ on the dual vacuum $\braC{\vac}$.

\begin{figure}
\begin{center}
\begin{tikzpicture}[scale=.3]
\draw (0,0) -- (1,0) -- (1,-5) -- (0,-5) -- (0,0);
\draw (0,-1) -- (1,-1);
\draw (0,-2) -- (1,-2);
\draw (0,-4) -- (1,-4);
\node at (.5,-2.5) {$\vdots$};
\draw (3,-5) -- (4,-5) -- (4,-10) -- (3,-10) -- (3,-5);
\draw (3,-6) -- (4,-6);
\draw (3,-7) -- (4,-7);
\draw (3,-9) -- (4,-9);
\node at (3.5,-7.5) {$\vdots$};
\node at (6,-10.5) {$\ddots$};
\draw (8,-11.5) -- (9,-11.5) -- (9,-16.5) -- (8,-16.5) -- (8,-11.5);
\draw (8,-12.5) -- (9,-12.5);
\draw (8,-13.5) -- (9,-13.5);
\draw (8,-15.5) -- (9,-15.5);
\node at (8.5,-14) {$\vdots$};
\end{tikzpicture}
\end{center}
\caption{General structure of disjoint strips}
\label{fig4}
\end{figure}
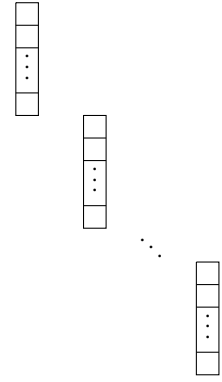

In order to formulate the vertical presentation of the action of the generators $W_{m,n}^C$ on this basis, we need to introduce again several new notations. Firstly, to a box $\Abox\in\l$ of coordinates $(i,j)$, we associate a new box content $\hchi_{\sAbox}=q^{(j-1)/2}$ depending only on the second coordinate $j$. Next, we introduce the notion of disjoint vertical strips that are sets of columns of boxes, the boxes coordinates $(i_k,j_k)$ satisfying $i_{k+1}\geq i_k$ and $j_{k+1}=j_k-1$ (see figure \ref{fig4}). We denote $A_m^C(\l)$ (resp. $R_m^C(\l)$) the set of disjoint strips $\rho$ with $2m$ boxes that can be added to (resp. removed from) $\l$, such that $\l\pm\rho\in$OP and without changing the number of columns (i.e. $\ell(\l\pm \rho)=\ell(\l)$). To such a strip we associate the content $\hchi_\rho=\hchi_{\sAbox}$ where $\Abox$ is the box in the top right corner (i.e. of maximal coordinate $j$). Thus, the boxes of a disjoint vertical strip have the contents $q^{-k/2}\hchi_\rho$ for $k=0,\cdots 2m-1$.

In addition, we introduce the set $A_m^{CP}(\l)$ (resp $R_m^{CP}(\l)$) of columns pairs with a total of $2m$ boxes that can be added to (resp. removed from) $\l$.\footnote{In fact $A_m^{CP}(\l)$ contains all pairs of column of odd heights with $2m$ boxes while $R_m^{CP}(\l)$ is the set of the pairs of columns of $\l$ containing $2m$ boxes. Note also that we need to count multiplicities: if $\l$ contains $k_i$ columns with the same height $\l_i$, and $k_j$ columns of height $\l_j$ with $\l_i+\l_j=2m$, $R_m^{CP}(\l)$ contains $k_ik_j$ pairs of columns $(\l_i,\l_j)$, or $k_i(k_i-1)/2$ pairs if $\l_i=\l_j$. The set $R_m^C(\l)$ is also degenerate: a strip has the multiplicity $\prod_ik_i$ if columns $\l_i$ to which the boxes are removed have multiplicity $k_i$.} To a pair of columns $\rho\in A_m^{CP}(\l)\cup R_m^{CP}(\l)$, we associate the content $\hchi_\rho=\hchi_{\sAbox}$ where $\Abox$ is the box at the top of the highest column (i.e. again of maximal coordinate $j$). We also associate the sign $(-1)^\rho=(-1)^{(k(\rho)-1)/2}$ where $k(\rho)$ is the height of the smallest column (the tallest column has the height $2m-k(\rho)$).

\begin{widetext}
With these notations, the vertical action of the generators reads for $m$ strictly positive
\begin{align}\label{WC_vert}
\begin{split}
W^C_{0,n}\ketC{\l}&=\left((1-q^{n/2})q^{-n/2}\sum_{\sAbox\in\l}(\hchi_{\sAbox}^n+q^{-n/2}\hchi_{\sAbox}^{-n})\right)\ketC{\l},\\
W^C_{m,n}\ketC{\l}&=\sum_{\rho\in R_m^C(\l)}\left((-1)^{m+1}q^{-mn}\hchi_\rho^n+q^{-n}\hchi_\rho^{-n}\right)\ketC{\l-\rho}+\sum_{\rho\in R_m^{CP}(\l)}(-1)^\rho\left((-1)^{m+1}q^{-mn}\hchi_\rho^n+q^{-n}\hchi_\rho^{-n}\right)\ketC{\l-\rho},\\
W^C_{-m,n}\ketC{\l}&=\sum_{\rho\in A_m^C(\l)}\left((-1)^{m+1}\hchi_\rho^n+q^{(m-1)n}\hchi_\rho^{-n}\right)\ketC{\l+\rho}-\!\!\!\!\sum_{\rho\in A_m^{CP}(\l)}\left(1-\hf\d_{k(\rho),m}\right)\left((-1)^{m+1}\hchi_\rho^n+q^{(m-1)n}\hchi_\rho^{-n}\right)(-1)^\rho\ketC{\l+\rho}.
\end{split}
\end{align}
\end{widetext}

The Kronecker $\d$ provides an extra factor $1/2$ when the two columns have the same size, $k(\rho)=m$. In particular, we have for the Heisenberg subalgebra with $k$ odd positive,
\begin{align}
\begin{split}
a_k\ketC{\l}&=2\sum_{\rho\in R_k^C(\l)}\ketC{\l-\rho}+2\sum_{\rho\in R_k^{CP}(\l)}(-1)^\rho\ketC{\l-\rho},\\
a_{-k}\ketC{\l}&=2\sum_{\rho\in A_k^C(\l)}\ketC{\l+\rho}\\
&-2\sum_{\rho\in A_k^{CP}(\l)}\left(1-\hf\d_{k(\rho),k}\right)(-1)^\rho\ketC{\l+\rho}.
\end{split}
\end{align}

The algebra $\CW^C$ can be supplemented with a grading operator $d_C$ such that $[d_C,W_{m,n}^C]=-2mW_{m,n}^C$. It is represented as
\begin{align}
\begin{split}
&d_C=\sum_{r\in\mZ+1/2}(-1)^{r-1/2}r:\hphi_{-r}\hphi_r:,\\
&d_C\ketC{\l}=|\l|\ketC{\l},\quad \a^{d_C} W_{m,n}^C\a^{-d_C}=\a^{-2m}.
\end{split}
\end{align}
Since the number of columns varies by $0,\pm 2$ under the action of the generators $W_{m,n}^C$, this representation is not irreducible but contains two sectors corresponding to partitions with either an odd or an even number of parts. In the following, we focus on the case of $\ell(\l)$ even.

\subsection{Symmetric polynomials}
In the case of $\hat C_\infty$, there are two ways of defining symmetric polynomials, depending on whether we generalize the bosonic expression \ref{def_bl}, or the fields expression \ref{bl_fields} of $\hat B_\infty$. In the first scenario, the fermionic component of the twisted bosonization is missing and, as a result, the q-difference equation cannot be established. The polynomials are those obtained by van de Leur, Orlov and Shiota in \cite{VanDeLeur2012} when they set their odd times to zero, and their half-integer times to $p_k(x)/k$.

In \cite{VanDeLeur2012}, the way around the problem of the missing fermionic component is the introduction of a supersymmetric vertex operator with both odd integer and half-integer times. Here, we will take a simpler approach that roughly corresponds to consider only a specific component of the supersymmetric amplitudes (up to the important question of normalization). This scenario corresponds to generalize the formula \ref{bl_fields} for the polynomials $b_\l^{(N)}(x)$ expressed as the correlator of a finite number of fields $N$ (the polynomial variables $x_i$ enter as the insertion positions). In this case, the q-difference equation associated to the diagonal action of $W_{0,n}^C$ can be derived. However, in contrast with the case of $\hat B_\infty$, the inductive limit that sends the number of variables $N$ to infinity is non-trivial here, and we were unable to define it properly. It would be interesting to derive a q-difference equation on the supersymmetric amplitudes that admit a proper large $N$ limit but it is beyond the scope of this paper.

\subsubsection{Bosonic approach and vOS polynomials}
Following van de Leur, Orlov and Shiota \cite{VanDeLeur2012}, we introduce the set of symmetric polynomials,
\begin{equation}\label{def_vOS}
C_\l(x)=\hf\braC{\vac}e^{\sum_{k>0}\dodd{k}\frac{p_k(x)}{k}a_k}\ketC{\l}.
\end{equation}
The normalization has been modified here, it is such that $C_{1^2}=p_1$. Then, e.g.
\begin{align}
\begin{split}
&C_{31}=p_1^2,\quad C_{1^4}=6p_1^2,\quad C_{51}=\dfrac23 p_1^3 + \dfrac13p_3,\quad C_{1^6}=60p_1^3,\\
&C_{3^2}=\dfrac43p_1^3 -\dfrac13p_3,\quad C_{71}=\dfrac13p_1^4 + \dfrac23p_1p_3,\\
&C_{3^21^2}=\dfrac{20}3p_1^4 - \dfrac23p_1p_3,\quad C_{1^8}=840p_1^4.
\end{split}
\end{align}
Since the number of columns of the Young diagram $\l$ varies by $0,\pm2$ under the action of $W_{k,0}^C$ on the state $\ketC{\l}$, the polynomials $C_\l(x)$ are non-zero only if $\ell(\l)$ (or $|\l|$ since $\l\in$OP) is even. However, more general (skew-type) polynomials can be introduced with
\begin{equation}
C_{\l/\mu}(x)=\hf\braC{\mu}e^{\sum_{k>0}\dodd{k}\frac{p_k(x)}{k}a_k}\ketC{\l}.
\end{equation}
These polynomials are non-zero when $\mu$ is a sub-diagram of $\l$ (i.e. $\mu\prec\l$) with $\ell(\l)-\ell(\mu)$ even. Inserting the grading operator $d_C$, one can show that these polynomials are homogeneous of degree $(|\l|-|\mu|)/2\in\mZ$. An explicit expression has been found for $C_\l(x)$ in \cite{VanDeLeur2012} as an Hafnian of Schur functions projected on odd power sums. The analogue of the Cauchy identity has also been obtained there, 
\begin{equation}
\sum_{\l\in\text{OP}}d_\l^{-2}C_\l(x)C_\l(y)=\prod_{i,j}\dfrac{1+x_iy_j}{1-x_iy_j},
\end{equation}
for certain normalization constant $d_\l$.

The derivation of Pieri-like rules follows from the same steps as before, namely the insertion of $W_{k,0}^C$ inside the correlator on the left, and the use of the Heisenberg algebra relations to move it to the right. We find \textbf{for $k$ odd}:
\begin{align}
\begin{split}\label{Pieri_C}
&p_k(x)C_\l(x)=-\sum_{\rho\in A_k^C(\l)}C_{\l+\rho}(x)\\
&\qquad\qquad+\sum_{\rho\in A_k^{CP}(\l)}\left(1-\hf\d_{k(\rho),k})\right)(-1)^\rho C_{\l+\rho}(x),\\
&k\dfrac{\p}{\p p _k(x)}C_\l(x)=2\sum_{\rho\in R_k^C(\l)}C_{\l-\rho}(x)+2\sum_{\rho\in R_k^{CP}(\l)}(-1)^\rho C_{\l-\rho}(x).
\end{split}
\end{align}
On the other hand, as anticipated, we were not able to derive a q-difference equation due to the missing fermionic operators in the correlator.

\subsubsection{From correlators to symmetric polynomials}
Instead of generalizing the bosonic formula, we can try to generalize the expression \ref{bl_fields} of the symmetric polynomials $b_\l(x)$ involving the Majorana fields. For this purpose, we need to consider a finite number of variables $N$. Moreover, in the absence of a bosonic normal-ordering, spoiled by the fermionic modes $\chi_r$, we need to introduce the notation
\begin{equation}
\vdots\hphi(z_1)\cdots\hphi(z_N)\vdots=\prod_{\superp{i,j=1}{i<j}}^N(z_i+z_j)\ \hphi(z_1)\cdots\hphi(z_N).
\end{equation} 
Since the prefactor kills the delta function arising in the r.h.s. when permuting the fields, the l.h.s. is invariant under the action of the symmetric group $\S_N$ on the coordinates $z_i$. In addition, it is easy to show the following properties\footnote{In order to obtain the expansion in powers of $z^{\mp1}$, we assumed that the operators are evaluated between two states $\braC{\l}\cdots\ketC{\mu}$. Then, the insertion of the field $\hphi(z)$ on the left will produce only a finite number of positive powers, while an insertion on the right produces a finite number of negative powers. The third property follows from the normal ordering $(z+w)\hphi(z)\hphi(w)=1+(z+w):\hphi(z)\hphi(w):$. Each matrix element $\braC{\l}:\hphi(z)\hphi(w):\ketC{\mu}$ has only a finite number of terms of order $z^kw^l$, and so the normal-ordered product is non-singular at $z+w=0$. Thus, the limit $z\to -w$ of the r.h.s. is simply one. Taking the difference of the first two properties, a delta function appears at each pole $z=-z_i$, and the residues simplify using the third property to give the final one.}
\begin{align}\label{propC}
\begin{split}
\text{I.}\quad &\hphi(z)\vdots\hphi(z_1)\cdots\hphi(z_N)\vdots=\left[\prod_{i=1}^N\dfrac1{z+z_i}\right]_+\vdots\hphi(z)\hphi(z_1)\cdots\hphi(z_N)\vdots,\\
\text{II.}\quad &\vdots\hphi(z_1)\cdots\hphi(z_N)\vdots\hphi(z)=\left[\prod_{i=1}^N\dfrac1{z+z_i}\right]_-\vdots\hphi(z)\hphi(z_1)\cdots\hphi(z_N)\vdots,\\
\text{III.}\quad &\lim_{z\to-z_i}\vdots\hphi(z)\hphi(z_1)\cdots\hphi(z_N)\vdots\\
&=\prod_{\superp{j=1}{j\neq i}}^N(z_j^2-z_i^2)\ \vdots\hphi(z_1)\cdots\cancel{\hphi(z_i)}\cdots\hphi(z_N)\vdots,\\
\text{IV.}\quad &[\hphi(z),\vdots\hphi(z_1)\cdots\hphi(z_N)\vdots],\\
&=\sum_{i=1}^Nz^{-1}\d(-z/z_i)\prod_{\superp{j=1}{j\neq i}}^N(z_i+z_j)\ \vdots\hphi(z_1)\cdots\cancel{\hphi(z_i)}\cdots\hphi(z_N)\vdots,
\end{split}
\end{align}
where $[f(z)]_\pm$ denotes the expansion of the function $f(z)$ in powers of $z^{\mp1}$.

We consider the following correlators that are symmetric functions of the $N$ variables $x_i$,
\begin{equation}\label{def_cNl}
c_\l^{(N)}(x)=\prod_{i=1}^Nx_i^{K_N}\ \braC{\vac}\vdots\hphi(-x_1^{-1})\cdots\hphi(-x_N^{-1})\vdots\ketC{\l},
\end{equation}
where $K_N$ is an integer to be determined soon. The r.h.s. vanishes unless $N+|\l|$ is even, and we assume that both $N$ and $|\l|$ are even for simplicity. As seen before, the quantities $c_\l^{(N)}(x)$ are symmetric functions of the variables $x_i$. Using the first property, we observe that they have a pole of order $x_1^{K_N-N+2}$ at $x_1=0$. Since there are no other singularities than the points at zero and infinity, and thanks to the $\S_N$-invariance, we conclude that the functions $c_\l^{(N)}(x)$ are symmetric polynomials if $K_N\geq N-2$. We choose the minimal value $K_N=N-2$. Then, using the grading operator $d_C$ and the fact that $\a^{d_C}\hphi(z)\a^{-d_C}=\a\hphi(\a^2 z)$, it is easy to show that the functions $c_\l^{(N)}(x)$ are homogeneous of degree $|\l|/2+N(N-2)/2$.

The simplest polynomial is associated to the vacuum, the corresponding correlator is computed in \cite{VanDeLeur2012} and takes the form of a Pfaffian,
\begin{equation}
c_\vac^{(N)}(x)=\prod_{\superp{i,j=1}{i<j}}^N\dfrac{(x_i+x_j)^2}{x_i-x_j}\ \text{Pf}\left(\dfrac{x_i-x_j}{(x_i+x_j)^2}\right).
\end{equation} 
To define the limit $N\to\infty$, we need the inductive property that $c_\l^{(N+2)}(x)$ reproduces $c_\l^{(N)}(x)$ when $x_{N+2}=x_{N+1}=0$. Using the previous properties, we can take this limit explicitly, and we find
\begin{equation}
\left.c_\l^{(N+2)}(x)\right|_{x_{N+2}=x_{N+1}=0}=\prod_{i=1}^Nx_i^2\ c_\l^{(N)}(x).
\end{equation} 
The extra factor $\prod_ix_i^2$ makes it difficult to take the limit $N\to\infty$. However, this factor is independent of $|\l|$ and thus the limit of the ratios $c_{\l}^{(N)}(x)/c_{\vac}^{(N)}(x)$ is well-defined. On the other hand, there is no indication that $c_{\vac}^{(N)}(x)$ divides $c_{\l}^{(N)}(x)$ and these ratios should be no longer polynomial (as opposed to the case of $b_\l^{(N)}(x)$).

We now turn to the action of the quantum W-algebra $\CW^C$. Using the bosonization rules, we have
\begin{align}
\begin{split}
c_\l^{(N)}(x)=&\prod_{i=1}^Nx_i^{-1}\prod_{\superp{i,j=1}{i<j}}^N\dfrac{(x_i+x_j)^2}{x_i-x_j}\times\\
&\braC{\vac}\chi(-x_1^{-1})\cdots\chi(-x_N^{-1})e^{\sum_{k>0}\dodd{k}\frac{p_k(x)}{k}a_k}\ketC{\l}.
\end{split}
\end{align}
Comparing with the expression \ref{def_vOS} of the polynomials $C_\l(x)$, we note that the major difference is the presence of the fermionic fields in the correlator. Since the modes $a_k$ commute with the fermionic fields, the polynomials $c_\l^{(N)}(x)$ obey the same rules \ref{Pieri_C} with respect to the multiplication by the elementary power sums $p_k(x)$. Note, however, that the action of the derivative $\p/\p p_k$ is different since the fermionic fields introduce an extra dependence on the coordinates $x_i$. The main reason for introducing these polynomials is that they satisfy the following q-difference equation associated to the diagonal action of the modes $W_{0,n}^C$ on the states $\ketC{\l}$,
\begin{align}
\begin{split}\label{C_qdiff}
&\sum_{i=1}^N\left(q^{-n/2}C_i^{(n)}(x)T_{q^{-n},x_i}-q^{n/2}C_i^{(-n)}(x)T_{q^n,x_i}\right)\ c_\l^{(N)}(x)\\
=&\left((1-q^{n/2})\sum_{\sAbox\in\l}(\hchi_{\sAbox}^n+q^{-n/2}\hchi_{\sAbox}^{-n})\right)c_\l^{(N)}(x),
\end{split}
\end{align}
with
\begin{equation}\label{def_Cin}
C_i^{(n)}(x)=\prod_{\superp{j=1}{j\neq i}}^N\dfrac{x_i+x_j}{q^{-n}x_i+x_j}
\end{equation} 
Once again, the derivation is given in appendix \ref{AppB}. Note that both sides of the equation are antisymmetric under the replacement $q\to q^{-1}$.

\section{$D$-type orbifold and $\oinf$}
The algebra $\hat D_\infty$, also called $\oinf$, is the central extension of the algebra of infinite matrices $d_{i,j}$ satisfying the relation $d_{i,j}+d_{1-i,1-j}=0$ \cite{FFZ}. It can be defined using the set of generators $\COD_{r,s}$ with half integer indices, and a central element $C$, satisfying the commutation relations
\begin{align}\label{def_oinf}
\begin{split}
[\COD_{r,s},\COD_{t,u}]&=\d_{s+t}\COD_{r,u}-\d_{r+u}\COD_{t,s}+2C(\th(r)-\th(t))\d_{s+t}\d_{r+u}\\
&+\d_{r+t}\COD_{u,s}-\d_{s+u}\COD_{r,t}+2C(\th(u)-\th(r))\d_{s+u}\d_{r+t},
\end{split}
\end{align}
together with the relations $\COD_{r,s}=-\COD_{s,r}$.  It is a subalgebra of $\glinf$ generated by the linear combinations $\COD_{r,s}=E_{r,s}-E_{s,r}$. The algebra $D_{\infty}$ is obtained using the representation of level zero defined in \ref{sec_def_rho}, it gives $\rho_0(\COD_{i-1/2,-j+1/2})=d_{i,j}$ with $d_{i,j}=e_{i,j}-e_{1-j,1-i}$.

The transformation \ref{rel_W_E} applied to the operators $\COD_{r,s}$ defines the generators
\begin{equation}\label{def_bhW}
W^D_{m,n}=\sum_{r\in\mZ+1/2}q^{-(r+1/2)n}\COD_{m-r,r}
\end{equation} 
of a subalgebra $\CW^D$ of $\CW$. Just like in the previous cases, this subalgebra is a $\mZ_2$-orbifold \cite{Lebedev1992}. It is obtained from the involutive automorphism
\begin{equation}\label{def_sD}
\s_D:\ W_{m,n}\to -q^{-(m+1)n}W_{m,-n},\quad C\to C.
\end{equation}
The subalgebra $\CW^D$ is generated by the central element $(1+\s_D)C=2C$ and the projections $W^D_{m,n}=(1+\s_D)W_{m,n}$, coinciding with the definitions \ref{def_bhW}. They satisfy the commutation relations
\begin{align}
\begin{split}\label{com_WD}
[W_{m,n}^D,W_{m',n'}^D]&=(q^{m'n}-q^{mn'})\left(W^D_{m+m',n+n'}+2C\dfrac{\d_{m+m'}}{1-q^{n+n'}}\right)\\
&-q^{-(m'+1)n'}(q^{m'n}-q^{-mn'})W^D_{m+m',n-n'}\\
&-q^{-(m'+1)n'}\dfrac{q^{m'n}-q^{-mn'}}{1-q^{n-n'}}2C\d_{m+m'}.
\end{split}
\end{align}
In this case, all the modes $W_{k,0}$ are projected out, and there is no Heisenberg subalgebra left. In order to define the horizontal presentation, we will need to introduce new operators. On the other hand, the modes $W_{0,n}^D$ are still commuting, they will be diagonal in the vertical presentation.

\subsection{Majorana representation}
The algebra $\oinf$ possesses a representation of level $C=1/2$ on the Fock space of a Majorana fermion with anti-periodic boundary conditions. This representation will be denoted $\rho^{(M)}$, it is defined upon the modes $\phi_r$ satisfying the anti-commutation relations $\{\phi_r,\phi_s\}=\d_{r+s}$ as $\rho^{(M)}(\COD_{r,s})=:\phi_r\phi_s:$ and 
\begin{equation}\label{WD_phi}
\rho^{(M)}(W_{m,n}^D)=\sum_{r\in\mZ+1/2}q^{-(r+1/2)n}:\phi_{m-r}\phi_r:.
\end{equation}
The Fock space is generated by the action of negative modes on the vacuum $\ketD{\vac}$ annihilated by positive modes, and the normal-ordering consists in moving positive modes to the right. Alternatively, the generators $W^D_{m,n}$ can be expressed as a contour integral of the fermionic field,\footnote{Note also
\begin{equation}
\{\phi(z),\phi(w)\}=z^{-1}\d(z/w),\quad \braD{\vac}\phi(z)\phi(w)\ketD{\vac}=\dfrac1{z-w}.
\end{equation}\protect\trick.}
\begin{align}
\begin{split}
&\rho^{(M)}(W_{m,n}^D)=\oint{\dfrac{dz}{2i\pi}z^m:\phi(z)\phi(q^nz):},\\
&\phi(z)=\sum_{r\in \mathbb{Z}+1/2} z^{-r-1/2}\phi_r.
\end{split}
\end{align}

\para{Decomposition of the Dirac representation} Since $\CW^D$ is a subalgebra of $\CW$, it can also be represented on the Dirac fermion Fock space. From equ. \ref{def_sD}, we find
\begin{equation}
\rho^{(D)}(W_{m,n}^D)=\oint{\dfrac{dz}{2i\pi}z^m\left(:\bpsi(z)\psi(q^nz):+:\psi(z)\bpsi(q^nz):\right)}.
\end{equation} 
This representation of $\CW^D$ has the level $\rho^{(D)}(C)=1$. It can be written as a sum of two Majorana representations with level $\rho^{(M)}(C)=1/2$ using the well known decomposition of a Dirac fermion into two Majorana fermions,
\begin{align}
\begin{split}\label{Dirac_Majorana}
&\psi(z)=\dfrac1{\sqrt{2}}\left(\phi^{(1)}(z)+i\phi^{(2)}(z)\right),\\
&\bpsi(z)=\dfrac1{\sqrt{2}}\left(\phi^{(1)}(z)-i\phi^{(2)}(z)\right).
\end{split}
\end{align}
As a result, we find for the generators
\begin{align}
\begin{split}
\rho^{(D)}(W_{m,n}^D)=\oint\dfrac{dz}{2i\pi}z^m&\Big(:\phi^{(1)}(z)\phi^{(1)}(q^nz):\\
&+:\phi^{(2)}(z)\phi^{(2)}(q^nz):\Big),
\end{split}
\end{align}
which implies that the Dirac representation decomposes as $\rho^{(D)}=\rho^{(M)}\otimes1+1\otimes\rho^{(M)}$.

\para{Horizontal presentation} In order to build a Heisenberg algebra, we need to consider a new set of operators acting on the Majorana Fock space, namely
\begin{align}
\begin{split}
\bW_{m,n}^D&=\oint{\dfrac{dz}{2i\pi}z^m:\phi(z)\phi(-q^nz):}\\
&=\sum_{r\in\mZ+1/2}(-1)^{r+1/2}q^{-(r+1/2)n}:\phi_{m-r}\phi_r:.
\end{split}
\end{align}
\begin{widetext}
These operators form a module for the adjoint action of $\CW^D$, they satisfy
\begin{align}
\begin{split}
&[W_{m,n}^D,\bW_{m',n'}^D]=(q^{m'n}-(-1)^mq^{mn'})\left(\bW^D_{m+m',n+n'}+\dfrac{\d_{m+m'}}{1+q^{n+n'}}\right)+(-1)^{m'}q^{-(m'+1)n'}(q^{m'n}-(-1)^{m}q^{-mn'})\left(\bW^D_{m+m',n-n'}+\dfrac{\d_{m+m'}}{1+q^{n-n'}}\right),\\
&[\bW_{m,n}^D,\bW_{m',n'}^D]=((-1)^{m'}q^{m'n}-(-1)^mq^{mn'})\left(W^D_{m+m',n+n'}+\dfrac{\d_{m+m'}}{1-q^{n+n'}}\right)+q^{-(m'+1)n'}(q^{m'n}-(-1)^{m+m'}q^{-mn'})\left(W^D_{m+m',n-n'}+\dfrac{\d_{m+m'}}{1-q^{n-n'}}\right).
\end{split}
\end{align}
\end{widetext}
In particular, the modes $a_k=\bW_{k,0}$ with $k$ even form the Heisenberg subalgebra $[a_k,a_l]=2k\d_{k+l}$. The operators act on the bosonic field as follows,
\begin{align}
\begin{split}\label{WD_com}
&[W_{m,n}^D,\phi(z)]=z^m\left(q^{-(m+1)n}\phi(q^{-n}z)-\phi(q^nz)\right),\\
&[\bW_{m,n}^D,\phi(z)]=z^m\left((-1)^{m+1}q^{-(m+1)n}\phi(-q^{-n}z)-\phi(-q^nz)\right),
\end{split}
\end{align}
and so $[a_k,\phi(z)]=-2z^k\phi(-z)$. Due to the sign flip of the field's argument, it is necessary to decompose into odd and even modes,
\begin{align}
\begin{split}
&\phi(z)=\phi_e(z)+z\phi_o(z),\\
&\phi_e(z)=\sum_{k\in\mZ}z^{-2k}\phi_{2k-1/2},\quad \phi_o(z)=\sum_{k\in\mZ}z^{-2k-2}\phi_{2k+1/2}.
\end{split}
\end{align}
Both field $\phi_e(z)$ and $\phi_o(z)$ are even functions of $z$. They transform as follows under the adjoint action of the Heisenberg modes,
\begin{equation}
[a_k,\phi_e(z)]=-2z^k\phi_e(z),\quad [a_k,\phi_o(z)]=2z^k\phi_o(z).
\end{equation} 
In addition, they satisfy the anticommutation relations
\begin{align}
\begin{split}
&\{\phi_e(z),\phi_e(w)\}=\{\phi_o(z),\phi_o(w)\}=0,\\
&\{\phi_e(z),\phi_o(w)\}=z^{-2}\d(z^2/w^2).
\end{split}
\end{align}
These relation are reproduced by the bosonization
\begin{align}
\begin{split}
&\phi_e(z)=e^QV_-(z)V_+(z)z^{2a_0},\quad \phi_o(z)=e^{-Q}V_-(z)^{-1}V_+(z)^{-1}z^{-2a_0},\\
&\text{with}\quad V_\pm(z)=e^{\pm\sum_{k>0}\d_{k,\text{even}}\frac{z^{\mp k}}{k}a_{\pm k}},\quad [a_0,Q]=1.
\end{split}
\end{align}
Defining the currents,
\begin{align}
\begin{split}
&w_n^D(z)=(1-q^n)\sum_{m\in\mZ}z^{-m}W_{m,n}^D+2C,\\
&\bar w_n^D(z)=(1+q^n)\sum_{m\in\mZ}z^{-m}\bar W_{m,n}^D+1,
\end{split}
\end{align}
they are represented on the fermionic Fock space as $(1-\s q^n)z\phi(z)\phi(\s q^nz)$ with $\s=1$ for $w_n(z)$ and $-1$ for $\bar w_n(z)$.
\begin{widetext}
This formula is bosonized into
\begin{align}
\begin{split}
&(1-\s q^n)(1-q^{2n})\sum_\pm q^{n(1\mp1)}z^{4\mp1}\ e^{\pm 2Q}\vdots e^{\pm \sum_{k\neq 0}\d_{k,\text{even}}\frac{z^{-k}}{k}(1+q^{-nk})a_k}\vdots z^{\pm 4a_0}q^{\pm 2na_0}+q^{n(1\mp1)/2}(1+\s q^n)\sum_\pm\vdots e^{\pm \sum_{k\neq 0}\d_{k,\text{even}}\frac{z^{-k}}{k}(1-q^{-nk})a_k}\vdots.
\end{split}
\end{align}
thus providing the horizontal presentation.

\end{widetext}

\subsection{Vertical presentation}\label{sec_Majorana}
The vertical presentation for $\hat D_\infty$ is similar to the one of $\hat B_\infty$ as it is possible to define a map $\phi_{\pm r}\to (\pm)^{r+1/2}\tphi_{\pm r\pm1/2}$ between the modes (although the zero-mode $\tphi_0$ must be considered separately). The states of the PBW basis can be parameterized either by strict partitions, or by symmetric partitions, using the bijection sending the columns height $\mu_i$ of the former to the hook lengths $h_i=2\mu_i-1$ of the latter. Here, we take the second stance and associate to a symmetric partition $\l$ with hook lengths $h_i\in2\mZ+1$, $i=1\cdots \ell(\l)$ the state\footnote{The similarity between these states and the Schur states in the Dirac representation can be understood by recalling that in the case of a Majorana fermion, particles and holes are identified and one may see them both as holes. The Schur states are defined as a product of operators $\bpsi_{-a_i-1/2}\psi_{-\ell_i-1/2}$ where the index $i$ runs over the boxes on the diagonal of $\l$, and $a_i$, $\ell_i$ are the arm and leg length resp. of the box of coordinate $(i,i)$. Formally, for a Majorana fermion, $\bpsi_{-a_i-1/2}\equiv\psi_{-\ell_i-1/2}\equiv\phi_{-h_i/2}$, and the equality of arm and leg lengths imposes a restriction to symmetric partitions.}
\begin{equation}\label{def_Majorana}
\ketD{\l}=\phi_{-h_1/2}\cdots \phi_{-h_{d(\l)}/2}\ketD{\vac}.
\end{equation}
Dual states $\braD{\mu}$ are defined using the adjoint action of the modes $\phi_r=\phi_{-r}^\dagger$ on the dual vacuum $\braD{\vac}$. It produces orthonormal states $\scalarD{\l}{\mu}=\d_{\l,\mu}$.

The action of the operators $W_{m,n}^D$ on the states $\ketD{\l}$ is obtained by a direct computation, using the commutation relations \ref{WD_com}. Fortunately, to express this action we can borrow most of the notations from the vertical presentation of $\hat B_\infty$. The result takes an unexpectedly compact form, for $m$ strictly positive
\begin{align}\label{WD_vert}
\begin{split}
&W^D_{0,n}\ketD{\l}=\left((1-q^n)q^{-n}\sum_{\sAbox\in\l}\chi_{\sAbox}^n\right)\ket{\l},\\
&W^D_{m,n}\ketD{\l}=\!\!\!\!\!\!\!\!\sum_{\rho\in R_m^B(\l)\cup R_m^{HP}(\l)}\!\!\!\!\!\!\!\!\left(q^{-mn}\chi_\rho^{-n}-q^{-n}\chi_\rho^n)\right)(-1)^{r(\rho)}\ketD{\l-\rho},\\
&W^D_{-m,n}\ketD{\l}=\!\!\!\!\!\!\!\!\sum_{\rho\in A_m^B(\l)\cup A_m^{HP}(\l)}\!\!\!\!\!\!\!\!\left(\chi_\rho^{-n}-q^{(m-1)n}\chi_\rho^n\right)(-1)^{r(\rho)}\ketD{\l+\rho}.
\end{split}
\end{align}
Generators $W^D_{m,n}$ with $m>0$ annihilates the vacuum, and $W^D_{m,0}\ketD{\l}=0$ for any $m$.

To obtain the action of the operators $\bW_{m,n}^D$, it is sufficient to replace $q^n$ by $-q^n$ in the action of $W_{m,n}^D$. It leads to introduce the extra signs $(-1)^{s(\sAbox)}=(-1)^{i-j}$ associated to a box $\Abox\in\l$. The sign $(-1)^{s(\rho)}$ of a strips $\rho\in A_m^B(\l)\cup R_m^B(\l)$ is again the sign of the box in the top left corner, and the sign of a hook pair $\rho$ is $(-1)^{s(\rho)}=(-1)^{k(\rho)-m+1}$. Then, we find
\begin{align}\label{bWD_vert}
\begin{split}
\bW^D_{0,n}\ketD{\l}&=\left(-(1+q^n)q^{-n}\sum_{\sAbox\in\l}(-1)^{s(\sAbox)}\chi_{\sAbox}^n\right)\ket{\l},\\
\bW^D_{m,n}\ketD{\l}&=\sum_{\rho\in R_m^B(\l)\cup R_m^{HP}(\l)}\left((-1)^{m}q^{-mn}\chi_\rho^{-n}+q^{-n}\chi_\rho^n)\right)\\
&\qquad\qquad(-1)^{r(\rho)+s(\rho)}\ketD{\l-\rho},\\
\bW^D_{-m,n}\ketD{\l}&=\sum_{\rho\in A_m^B(\l)\cup A_m^{HP}(\l)}\left(\chi_\rho^{-n}+(-1)^mq^{(m-1)n}\chi_\rho^n\right)\\
&\qquad\qquad(-1)^{r(\rho)+s(\rho)}\ketD{\l+\rho}.
\end{split}
\end{align}
In particular, for $k$ even positive,
\begin{align}\label{ak_Majorana}
\begin{split}
&a_k\ketD{\l}=2\sum_{\rho\in R_k^B(\l)\cup R_k^{HP}(\l)}(-1)^{r(\rho)+s(\rho)}\ketD{\l-\rho},\\
&a_{-k}\ketD{\l}=2\sum_{\rho\in A_k^B(\l)\cup A_k^{HP}(\l)}(-1)^{r(\rho)+s(\rho)}\ketD{\l+\rho}.
\end{split}
\end{align}

As in the other cases, the algebra $\CW^D$ can be supplemented by a grading operator $d_D$ such that $[d_D,W_{m,n}^D]=-2mW_{m,n}^D$. It is represented as
\begin{align}
\begin{split}
&d_D=\sum_{r\in\mZ+1/2}r:\phi_{-r}\phi_r:,\\
&d_D\ketD{\l}=|\l|\ketD{\l},\quad \a^{d_D} W_{m,n}^D\a^{-d_D}=\a^{-2m}.
\end{split}
\end{align}
It is also possible to show that $[d_D,\bW_{m,n}^D]=-2m\bW_{m,n}^D$, or  $\a^{d_D}\bW_{m,n}^D\a^{-d_D}=\a^{-2m}$. The number of hooks varies by $0,\pm 2$ under the action of the generators $W_{m,n}^D$ (and the operators $\bW_{m,n}^D$). Thus, the representation is not irreducible but contains two sectors corresponding to partitions with either an odd or an even number of hooks.


\subsection{Symmetric polynomials}
The discussion in this subsection is parallel to the case of $\hat C_\infty$, with the possibility of constructing two types of polynomials depending on whether we generalize the bosonic expression \ref{def_bl} or the expression \ref{bl_fields} as a correlator of Majorana fields. Thus, the symmetric polynomials $D_\l(x)$, and the skew-polynomials $D_{\l/\mu}(x)$ are defined as
\begin{align}
\begin{split}
&D_\l(x)=\braD{\vac}e^{\sum_{k>0}\deven{k}\frac{p_k(x)}{k}a_k}\ketD{\l},\\
&D_{\l/\mu}(x)=\braD{\mu}e^{\sum_{k>0}\deven{k}\frac{p_k(x)}{k}a_k}\ketD{\l}.
\end{split}
\end{align}
Since the action of $a_k$ with $k>0$ on a state $\ketD{\l}$ produces a finite sum of states $\ketD{\mu}$ with $\mu\prec\l$, these polynomials decompose into a finite sum of products of elementary power sums. Inserting the grading operator $d_D$, it is easily shown that they are homogeneous polynomials of degree $|\l|/2$ and $(|\l|-|\mu|)/2$ respectively. The derivation of Pieri-like rules follows from the same technique as before and we have for $k$ even
\begin{align}
\begin{split}
&p_k(x)D_\l(x)=\sum_{\rho\in A_k^B(\l)\cup A_k^{HP}(\l)}(-1)^{r(\rho)+s(\rho)}D_{\l+\rho}(x),\\
&k\dfrac{\p}{\p p _k(x)}D_\l(x)=2\sum_{\rho\in R_k^B(\l)\cup R_k^{HP}(\l)}(-1)^{r(\rho)+s(\rho)}D_{\l-\rho}(x).
\end{split}
\end{align}
A priori, these polynomials do not satisfy any q-difference equation that could be related to the action of either operator $W_{0,n}^D$ or $\bW_{0,n}^D$.

The definition of the polynomials $c_\l^{(N)}(x)$ with $N$ variables also extends to the case of $\hat D_\infty$. For this purpose, we introduce the notation
\begin{equation}
\vdots\phi(z_1)\cdots\phi(z_N)\vdots=\prod_{\superp{i,j=1}{i<j}}^N(z_i-z_j)\ \phi(z_1)\cdots\phi(z_N),
\end{equation} 
that will play the role of a bosonic normal ordering. This quantity is invariant under the action of the symmetric group $\S_N$ on the coordinates $z_i$, and obeys the following properties
\begin{align}
\begin{split}
\text{I.}\quad&\phi(z)\vdots\phi(z_1)\cdots\phi(z_N)\vdots=\left[\prod_{i=1}^N\dfrac1{z-z_i}\right]_+\vdots\phi(z)\phi(z_1)\cdots\phi(z_N)\vdots,\\
\text{II.}\quad&\vdots\phi(z_1)\cdots\phi(z_N)\vdots\phi(z)\\
&=(-1)^N\left[\prod_{i=1}^N\dfrac1{z-z_i}\right]_-\vdots\phi(z)\phi(z_1)\cdots\phi(z_N)\vdots,\\
\text{III.}\quad&\lim_{z\to z_i}\vdots\phi(z)\phi(z_1)\cdots\phi(z_N)\vdots\\
&=\prod_{\superp{j=1}{j\neq i}}^N(z_i-z_j)^2\ \vdots\phi(z_1)\cdots\cancel{\phi(z_i)}\cdots\phi(z_N)\vdots,\\
\text{IV.}\quad&[\phi(z),\vdots\phi(z_1)\cdots\phi(z_N)\vdots\}_{(-1)^{N+1}}\\
&=\sum_{i=1}^Nz^{-1}\d(z/z_i)\prod_{\superp{j=1}{j\neq i}}^N(z_i-z_j)\ \vdots\phi(z_1)\cdots\cancel{\phi(z_i)}\cdots\phi(z_N)\vdots,\\
\end{split}
\end{align}
with $[A,B\}_\pm$ coinciding with an anticommutator for the plus sign, and a commutator for the minus sign.

We consider the following correlators that are symmetric functions of the $N$ variables $x_i$,
\begin{equation}
d_\l^{(N)}(x)=\prod_{i=1}^Nx_i^{N-2}\ \braD{\vac}\vdots\phi(x_1^{-1})\cdots\phi(x_N^{-1})\vdots\ketD{\l}.
\end{equation}
Notice that the r.h.s. vanishes unless $N+d(\l)$ is even, and we will assume that both $N$ and $d(\l)$ are even. Just like $c_\l^{(N)}(x)$, the correlators $d_\l^{(N)}(x)$ are symmetric polynomials of the variables $x_i$. Using the grading operator, one shows that they are homogeneous of degree $|\l|/2+N(N-2)/2$. Unfortunately, the definition of an inductive limit suffers from the same problem since as $x_{N+2},x_{N+1}\to 0$,
\begin{equation}
\left.d_\l^{(N+2)}(x)\right|_{x_{N+2}=x_{N+1}=0}=\prod_{i=1}^Nx_i^{2}\ d_\l^{(N)}(x).
\end{equation}
The simplest polynomial can be evaluated as a Pfaffian,
\begin{align}
\begin{split}
d_\vac^{(N)}(x)&=\prod_{\superp{i,j=1}{i<j}}^N(x_j-x_i)\ \text{Pf}\left(\dfrac1{x_j-x_i}\right).
\end{split}
\end{align}
Up to a gaussian factor, this polynomial coincides with the (bosonic) Moore-Read wavefunction introduced in the context of the fractional quantum Hall effect \cite{Moore1991}, and can be expressed as a Jack polynomial $J_\l^{\a}(x)$ with the negative coupling $\a=-3$ \cite{Bernevig2008}. It would be interesting to see if the other polynomials $d_\l^{(N)}$ also play a role in this context.\footnote{We would like to thank the anonymous referee for this important remark.}

We now turn to the action of the quantum W-algebra $\CW^D$. Due to the extra sign in the argument of the field for the commutator $[a_k,\phi(z)]=-2z^k\phi(-z)$, in place of the Pieri rules we find the weird looking relation
\begin{align}
\begin{split}
&\sum_{i=1}^Nx_i^k\prod_{\superp{j=1}{j\neq i}}^N\dfrac{x_j-x_i}{x_j+x_i}T_{-1,x_i}\ d_\l^{(N)}(x)\\
=&-\sum_{\rho\in A_k^B(\l)\cup A_k^{HP}(\l)}(-1)^{r(\rho)+s(\rho)}d_{\l+\rho}^{(N)}(x).
\end{split}
\end{align}
Finally, the q-difference equation is obtained in exactly the same way as in the case of $\hat C_\infty$, and it didn't seem necessary to include the derivation. It reads
\begin{align}
\begin{split}\label{D_qdiff}
&\left(\sum_{i=1}^Nq^{n/2}D_i^{(n)}(x)T_{q^{-n},x_i}-q^{-n/2}D_i^{(-n)}(x)T_{q^{n},x_i}\right)d_\l^{(N)}(x)\\
=&-(q^{n/2}-q^{-n/2})\sum_{\sAbox\in\l}\chi_\sAbox^n\ d_\l^{(N)}(x),
\end{split}
\end{align}
with
\begin{equation}
D_i^{(n)}(x)=\prod_{\superp{j=1}{j\neq i}}^N\dfrac{x_i-x_j}{q^{-n}x_i-x_j}
\end{equation}
The same equation with $q^n$ replaced by $-q^n$ is obtained from the insertion of $\bW_{0,n}^D$ instead.

\section{Concluding remarks}
The main results presented in this paper are the vertical presentations \ref{WB_vert}, \ref{WC_vert} and \ref{WD_vert} for the subalgebras $\CW^X$ of quantum $\Winf$ with $X=B,C,D$. These formulas express the action of the generators on the Fock states in a combinatorial manner. They were further used to derive a set of rules and a q-difference equation for a corresponding set of symmetric polynomials. In the case of $\CW^B$, these polynomials are Q-Schur functions indexed by strict partitions. Since they can be obtained as a specialization of Hall-Littlewood polynomials at $t=-1$, it is natural to wonder whether Hall-Littlewood polynomials would satisfy a $t$-deformed q-difference equation. Of course, this equation has to be different from the $q=0$ limit of the Macdonald equation they already satisfy. Furthermore, the expressions \ref{WB_phi}, \ref{WC_phi} and \ref{WD_phi} of the generators $W_{m,n}^X$ in terms of fermionic/bosonic modes appears very generic. It is tempting to use the same expression with Jing's $t$-fermions \cite{Jing1991} to define a quantum W-algebra. It would be interesting to investigate further the possible connections between this algebra, the Hall-Littlewood polynomials and the quantum affine algebra of $\mathfrak{sl}(2)$ obtained in the $q\to0$ limit of the quantum toroidal $\mathfrak{gl}(1)$ algebra.

In the case of $\CW^C$ and $\CW^D$, we were led to introduce two sets of symmetric polynomials for each algebra. The first set, denoted $C_\l(x)$ and $D_\l(x)$ is built using the action of an Heisenberg algebra on the Fock states of the representation. The polynomials can be define for infinitely many variables, they obey Pieri-like rules under the multiplication by power-sums. However, they do not satisfy any q-difference equation (a priori). The second sets of polynomials, denoted $c_\l^{(N)}(x)$, $d_\l^{(N)}(x)$, are defined with finitely many variables $N$ and do obey the q-difference equations \ref{C_qdiff} and \ref{D_qdiff} that naturally extend the equation \ref{B_qdiff} obtained for $\CW^B$. In the case of $\CW^C$, the polynomials $C_\l(x)$ and $c_\l^{(N)}(x)$ can be related using a formalism based on supersymmetric variables introduced in \cite{VanDeLeur2012}. We expect that a similar formalism exists in the case of $\CW^D$ as well. Then, it may be possible to introduce a supersymmetric version of the algebra $\CW^X$ that would act on the correlators of the superfields.

Our main motivation for the study of these subalgebras is the application to topological string theory. Using both vertical and horizontal presentations, one may be able to define the equivalent of the Awata-Feigin-Shiraishi (AFS) intertwiner that provides the operator form of the refined topological vertex \cite{AFS}. In particular, our hope is that the $\CW^B$ algebra would provide the quantum algebraic structure missing in the earlier attempts by Foda and Wheeler to introduce a B-type topological vertex. The latter were based upon the Okounkov-Reshetikhin-Vafa melting crystal picture \cite{ORV} and used B-type plane partitions \cite{FW} and Jing's $t$-fermions \cite{Foda2008}.

Going in a different direction, one could try instead to extend the relation between topological strings and integrable hierarchies \cite{Aganagic2003}. In \cite{Nakatsu2007,Takasaki2018}, Nakatsu and Takasaki have shown that a certain time-deformation of a topological string amplitude is a tau function of the KP hierarchy. \footnote{In fact, this property follows from the intertwining relation obeyed by the vacuum component of the AFS intertwiner (see \cite{Bourgine2021}). Thus, it would seem that we are uncovering different aspects of a bigger picture.}It would be instructive to extend their calculation to the 
BKP, CKP and DKP hierarchies. For this purpose, understanding the connections with symmetric polynomials appears to be an essential ingredient.

Finally, the toroidal Yangian of $\gl(1)$, which is obtained from the quantum toroidal $\gl(1)$ algebra in the degenerate limit $q_1,q_2\to1$, provides the COhomological Hall Algebra of the ADHM quiver \cite{Nakajima1994,Nakajima1999}. It would be interesting to relate the $q\to1$ limit of the $\CW^X$ subalgebras to the cohomology of a quiver variety.

\section*{Acknowledgments}
The paper is dedicated to the memory of Omar Foda. The author's interest in the algebraic structures presented in this article arose from discussions with him at the University of Melbourne. Omar had an impressive intuition on this topic, his unique way of thinking and his invaluable advice are sorely missed, as much as his warm friendship. The author would like to thank also Sasha Garbali, Yutaka Matsuo, Vincent Pasquier and Kilar Zhang for very helpful discussions. He is also very grateful to Pr. Kimyeong Lee and the Korea Institute for Advanced Study (KIAS) for their generous support in these difficult times. This research was partly supported by the Basic Science Research Program through the National Research Foundation of Korea (NRF) funded by the Ministry of Education through the Center for Quantum Spacetime (CQUeST) of Sogang University (NRF-2020R1A6A1A03047877).

\textit{Data sharing not applicable – no new data generated}

\appendix

\section{Bosonic correlators in Schur basis}\label{AppA}
It is an established fact that skew-Schur polynomials can be written as correlators in the free boson Fock space,
\begin{align}
\begin{split}\label{slm_Gamma}
&s_{\l/\mu}(x)=\bra{\l}\G_-(x)\ket{\mu}=\bra{\mu}\G_+(x)\ket{\l},\\
&\Gamma_\pm(x) =\exp\left(\sum_{k>0} \frac{p_k(x)}{k}a_{\pm k} \right).
\end{split}
\end{align}
In the particular case $\mu=\vac$, we recover the formula \ref{Schur_corr}. In this appendix, we use the correspondence between symmetric polynomials and the free boson Fock space to derive a proof of this formula from well-known identities obeyed by skew-Schur polynomials. Recall that the Schur polynomials $s_\l(x)$ correspond to the bosonic states $\ket{\l}_x$ under this correspondence, where we indicate by the label $x$ the states belonging to the Fock space associated to polynomials in variables $x$ (we will be considering several Fock spaces).

As a warm-up, we consider the following formula from \cite{Macdonald},
\begin{equation}\label{Mac1}
\sum_\l s_\l(x)s_\l(y)=\prod_{i,j}(1-x_iy_j)^{-1}=\exp\left(\sum_{k>0}\dfrac1kp_k(x)p_k(y)\right).
\end{equation} 
Taking the correspondence along the $y$ Fock space, we have
\begin{align}
\begin{split}
&\sum_\l s_\l(x)\ket{\l}_y=\exp\left(\sum_{k>0}\dfrac1kp_k(x)a_{-k}\right)\ket{\vac}_y=\G_-(x)\ket{\vac}_y\\
&\implies s_\l(x)=\bra{\l}\G_-(x)\ket{\vac}.
\end{split}
\end{align}

\begin{widetext}
Next, we consider the formula \cite{Macdonald} (page 71)
\begin{align}
\begin{split}\label{Mac71}
\sum_{\l,\mu}s_{\l/\mu}(x)s_\l(y)s_\mu(z)&=\prod_{i,j}(1-x_iy_j)^{-1}\times\prod_{i,j}(1-y_iz_j)^{-1}=\exp\left(\sum_{k>0}\dfrac1k(p_k(x)p_k(y)+p_k(y)p_k(z))\right).
\end{split}
\end{align}
Taking the correspondence for the variables $y$ and $z$, we get
\begin{equation}
\sum_{\l,\mu}s_{\l/\mu}(x)\ \ket{\l}_y\otimes\ket{\mu}_z=\exp\left(\sum_{k>0}\dfrac1k(p_k(x)a_{-k}\otimes1+a_{-k}\otimes a_{-k})\right)\ \ket{\vac}_y\otimes\ket{\vac}_z.
\end{equation} 
Inserting the identity in the $y$-space, and projecting on $ _{\ y\!\!}\bra{\l}\otimes_{\ z\!\!}\bra{\mu}$, we find
\begin{equation}\label{slm}
s_{\l/\mu}(x)=\sum_{\eta}\ _{\ y\!\!}\bra{\l}e^{\sum_{k>0}\frac1kp_k(x)a_{-k}}\ket{\eta}_y\left(_{\ y\!\!}\bra{\eta}\otimes_{\ z\!\!}\bra{\mu} e^{\sum_{k>0}\frac1k a_{-k}\otimes a_{-k}}\ \ket{\vac}_y\otimes\ket{\vac}_z\right).
\end{equation} 
On the other hand, the identity \ref{Mac1} also implies
\begin{equation}
e^{\sum_{k>0}\frac1k a_{-k}\otimes a_{-k}}\ \ket{\vac}_y\otimes\ket{\vac}_z=\sum_{\l}\ket{\l}_y\otimes\ket{\l}_z, 
\end{equation} 
and, as a result, a delta function $\d_{\eta,\mu}$ appears in the RHS of \ref{slm}, leading to
\begin{equation}
s_{\l/\mu}(x)=\bra{\l}e^{\sum_{k>0}\frac1kp_k(x)a_{-k}}\ket{\mu}=\bra{\l}\G_-(x)\ket{\mu}.
\end{equation}
Moreover, since $\G_+(x)=\G_-(x)^\dagger$, the second equality also holds. It is worth mentioning that starting instead from the relations\footnote{The first relation is given in \cite{Macdonald} (page 65) while the second one can be derived from \ref{Mac71} using the homomorphism \cite{Macdonald} (page 42) $\o$ on the $x$ variables. The latter acts as $\o(s_{\l/\mu})=s_{\l'/\mu'}$ on skew-Schur polynomials and $\o(p_k)=(-1)^{k-1}p_k$ on elementary power sums.}
\begin{align}
\begin{split}
&\sum_\l s_\l(x)s_{\l'}(y)=\prod_{i,j}(1+x_iy_j)=\exp\left(-\sum_k\dfrac1kp_k(-x)p_k(y)\right),\\
&\sum_{\l,\mu}s_{\l/\mu}(x)s_{\l'}(y)s_{\mu'}(z)=\prod_{i,j}(1+x_iy_j)\times\prod_{i,j}(1-y_iz_j)^{-1}=\exp\left(\sum_{k>0}\dfrac1k(-p_k(-x)p_k(y)+p_k(y)p_k(z))\right).
\end{split}
\end{align}
we can show similar formulas involving the transposed Young diagrams $\l'$,
\begin{equation}\label{slm_Gamma_2}
s_{\l/\mu}(x)=\bra{\l'}\G_-(-x)^{-1}\ket{\mu'}=\bra{\mu'}\G_+(-x)^{-1}\ket{\l'},
\end{equation} 
\end{widetext}

\section{Derivation of the q-difference equations}\label{AppB}
\subsection{Schur case}
We derive in this appendix the q-difference equation \ref{qdiff_Schur} obeyed by the Schur polynomials from the diagonal action of the modes $W_{0,n}$ in the vertical presentation. Inserting these operators in the r.h.s. of equ. \ref{Schur_corr}, and using \ref{Wmn_Fock} we find
\begin{equation}\label{insert_W0n}
\bra{\vac}e^{\sum_{k>0}\frac{p_k(x)}{k}a_{k}}W_{0,n}\ket{\l}=\left((1-q^{n})q^{-n}\sum_{\sAbox\in\l}\chi_{\sAbox}^n\right)s_\l(x).
\end{equation}
On the other hand, the operators $W_{0,n}$ are the zero modes of the currents $w_n(z)$ defined in \ref{def_wn}.
\begin{widetext}
Using the bosonized version of the current, we find after normal-ordering that
\begin{equation}
e^{\sum_{k>0}\frac{p_k(x)}{k}a_{k}}\left((1-q^n)W_{0,n}+1\right)=\oint_0\dfrac{dz}{2i\pi z} \prod_{i}\dfrac{1-q^nx_iz}{1-x_iz}\vdots e^{\sum_{k>0}\frac{p_k(x)}{k}a_k}w_n(z)\vdots.
\end{equation} 
The contour of the integral can be deformed on the sphere, thus picking up residues at $z=x_i^{-1}$ and infinity, and we find
\begin{align}
\begin{split}
&(1-q^n)\sum_{i=1}\prod_{j\neq i}\dfrac{1-q^nx_j/x_i}{1-x_j/x_i}e^{\sum_{k>0}x_i^{-k}(1-q^{nk})\frac{a_{-k}}{k}}e^{\sum_{k>0}\left(p_k(x)+x_i^k(q^{-nk}-1)\right)\frac{a_k}{k}}-\oint_\infty\dfrac{dz}{2i\pi z} \prod_{i}\dfrac{1-q^nx_iz}{1-x_iz}\vdots e^{\sum_{k>0}\frac{p_k(x)}{k}a_k}w_n(z)\vdots.
\end{split}
\end{align}
The negative modes disappear after projection on $\bra{\vac}$, and the contribution at infinity simplifies drastically since the operator has no longer any positive power of $z$. As a result, we get
\begin{equation}
\bra{\vac}e^{\sum_{k>0}\frac{p_k}{k}a_{k}}\left(W_{0,n}+\dfrac{1-q^{nN}}{1-q^n}\right)=\sum_{i=1}^N\prod_{j\neq i}\dfrac{x_i-q^nx_j}{x_i-x_j}\bra{\vac}e^{\sum_{k>0}\left(p_k(x)+x_i^k(q^{-nk}-1)\right)\frac{a_k}{k}},
\end{equation}
where $N$ is the number of variables which is used as a cut-off. The shift of $p_k(x)$ in the exponent of the $i$th term corresponds to the replacement of the variable $x_i$ by $q^{-n}x_i$, this operation is denoted $T_{q^{-n},x_i}$. Projecting on the state $\ket{\l}$ and combining the result with the relation \ref{insert_W0n}, we establish the q-difference equation \ref{qdiff_Schur}.
\end{widetext}

\para{Power sum $p_1$} The simplest (non-trivial) Schur polynomial is the one associated to the partition $1$ containing a single box, it is simply the first symmetric power sum $p_1(x)$. This polynomial should satisfy the q-difference equation \ref{qdiff_Schur}, which reads in this case
\begin{equation}\label{B4}
\left(q^n\sum_{i=1}^N\prod_{j\neq i}\dfrac{x_i-q^nx_j}{x_i-x_j}T_{q^{-n},x_i}-\sum_{i=1}^N q^{ni}\right)p_1(x)=(1-q^n)p_1(x).
\end{equation} 
It is instructive to check this equation against a direct calculation. Using $T_{q^{-n},x_i}p_1(x)=(q^{-n}-1)x_i+p_1(x)$, it can be rewritten in the form
\begin{align}
\begin{split}\label{B5}
&\left(S_1(x)+\dfrac{q^n}{1-q^n}S_0(x)p_1(x)-q^n\dfrac{1-q^{nN}}{(1-q^n)^2}\right)=p_1(x),\\
&\text{with}\quad S_k(x)=\sum_{i=1}^Nx_i^k\prod_{j\neq i}\dfrac{x_i-q^nx_j}{x_i-x_j}.
\end{split}
\end{align}
The sums $S_k(x)$ can be computed using Jacobi identities, but we prefer to present here an alternative derivation. We introduce the function
\begin{equation}
S(z)=\prod_i\dfrac{z-q^nx_i}{z-x_i}=1+(1-q^n)\sum_i\dfrac{x_i}{z-x_i}\prod_{j\neq i}\dfrac{x_i-q^nx_j}{x_i-x_j},
\end{equation} 
where the second equality is found by decomposition over the poles. Considering the value $S(0)$ and the subleading term in the large $z$ asymptotic expansion, we can show that
\begin{equation}
S_0(x)=\dfrac{1-q^{nN}}{1-q^n},\quad S_1(x)=p_1(x).
\end{equation}
Plugging this back into the l.h.s. of the equation \ref{B5} we find $p_1$, which shows that the first power sum does indeed obey the q-difference equation \ref{B4}.

\subsection{B-case}
The derivation of the q-difference equation for $b_\l(x)$ is very similar to the Schur case and uses the bosonization of the Majorana fermion. We first introduce the equivalent of the currents $w_n(z)$ for the subalgebra $\CW^B$,
\begin{equation}\label{def_wnB}
w_n^B(z)=(1-q^n)\sum_{m\in\mZ}z^{-m}W_{m,n}^B+2C.
\end{equation} 
In the Majorana representation, they read $w_n^B(z)=(1-q^n)\tphi(z)\tphi(-q^nz)$ which is bosonized into
\begin{equation}\label{wnB_boson}
w_n^B(z)=\hf(1+q^n)\vdots e^{-\sum_{k\in\mZ}\dodd{k}\frac{z^{-k}}{k}(1-q^{-nk})a_k}\vdots.
\end{equation}
We also introduce the shortcut notation
\begin{align}
\begin{split}\label{def_GB}
&\G_B(x)=e^{\sum_{k>0}\dodd{k}\frac{p_k(x)}{k}W_{k,0}^B}\\
\implies &b_\l(x)=\braB{\vac}\G_B(x)\ketB{\l},\quad b_\l^\ast(x)=2\braB{\Abox}\G_B(x)\ketB{\l}.
\end{split}
\end{align}

\begin{widetext}
After these preliminaries, we consider the insertion of the operator $W^B_{0,n}$ in the correlator \ref{def_bl} and use its diagonal action \ref{WB_vert} to show that 
\begin{align}
\begin{split}\label{WB0n_action}
&\braB{\vac}\G_B(x)W^B_{0,n}\ketB{\l}=\left((1-q^n)\sum_{\sAbox\in\l_-}(q^{-n}\chi_{\sAbox}^n+\chi_{\sAbox}^{-n})-\hf \right)\ b_\l(x).
\end{split}
\end{align}
To derive the q-difference equation, we use the fact that the operators $W_{0,n}^B$ correspond to the zero-modes of the current $w_n^B(z)$. From the bosonic expressions \ref{wnB_boson} and \ref{def_GB}, we compute the normal-ordering 
\begin{align}
\begin{split}
&\oint_0\dfrac{dz}{2i\pi z}\braB{\vac}\G_B(x)w_n^B(z)\ketB{\l}=\hf(1+q^n)\oint_0\dfrac{dz}{2i\pi z}\prod_i\dfrac{(1+x_i z)(1-q^nx_iz)}{(1-x_i z)(1+q^nx_iz)}\braB{\vac}\vdots e^{\sum_{k>0}\frac{\dodd{k}}{k}(p_k(x)-z^{-k}+q^{-nk}z^{-k})a_k}\vdots\ketB{\l}.
\end{split}
\end{align}
Negative modes were removed from the correlator since they annihilate the dual vacuum $\bra{\vac}$. Deforming the contour on the sphere, we pick up residues at $z=x_i^{-1},-q^{-n}x_i^{-1}$ and infinity,
\begin{align}
\begin{split}
\hf(1+q^n)\braB{\vac}\vdots e^{\sum_{k>0}\frac{\dodd{k}}{k}p_k(x)a_k}\vdots\ketB{\l}+(1-q^n)\sum_iB_i^{(n)}(x)\braB{\vac}\vdots e^{\sum_{k>0}\frac{\dodd{k}}{k}(p_k(x)-x_i^{k}+q^{-nk}x_i^{k})a_k}\vdots\ketB{\l}\\
-(1-q^n)\sum_iB_i^{(-n)}(x)\braB{\vac}\vdots e^{\sum_{k>0}\frac{\dodd{k}}{k}(p_k(x)+q^{nk}x_i^{k}-x_i^{k})a_k}\vdots\ketB{\l}
\end{split}
\end{align}
with the residues $B_i^{(n)}(x)$ given in \ref{def_Ai}. This expression can be rewritten as the action of the q-difference operators on the symmetric polynomials
\begin{equation}
\oint_0\dfrac{dz}{2i\pi z}\braB{\vac}\G_B(x)w_n^B(z)\ketB{\l}=(1-q^n)\left[\sum_i\left(B_i^{(n)}(x)T_{q^{-n},x_i}-B_i^{(-n)}(x)T_{q^n,x_i}\right)+\hf\dfrac{1+q^n}{1-q^n}\right]b_\l(x).
\end{equation} 
Combining this relation with \ref{WB0n_action}, we find the q-difference equation \ref{B_qdiff}. The derivation extends to $b_\l^\ast(x)$ since the property  $\braB{\Abox}a_k=0$ for $k<0$ also holds.  
\end{widetext}

\para{Power sum $p_1$} 
For $b_{2^2}=p_1$, the q-difference equation \ref{B_qdiff} reads
\begin{equation}\label{qdiff_p1}
\sum_i\left(B_i^{(n)}(x)T_{q^{-n},x_i}-B_i^{(-n)}(x)T_{q^n,x_i}\right)\ p_1(x)=q^{-n}(1-q^{2n})\ p_1(x).
\end{equation} 
We can check this formula against a direct computation. We have
\begin{align}
\begin{split}\label{equ_p1}
&\left(\sum_i\left(B_i^{(n)}(x)T_{q^{-n},x_i}-B_i^{(-n)}(x)T_{q^{n},x_i}\right)\right)p_1(x)\\
=&-(1-q^{-n})S_1^B(x)+S_0^B(x)p_1(x),\\
\text{with}\quad &S_0^B(x)=\sum_i\left(B_i^{(n)}(x)-B_i^{(-n)}(x)\right),\\
&S_1^B(x)=\sum_i\left(x_iB_i^{(n)}(x)+q^nx_iB_i^{(-n)}(x)\right).
\end{split}
\end{align}
In order to compute the sums $S_k^B(x)$. we consider the function
\begin{align}
\begin{split}
B(z)&=\prod_i\dfrac{(z+x_i)(z-q^nx_i)}{(z-x_i)(z+q^nx_i)}\\
&=1+2\dfrac{1-q^n}{1+q^n}\sum_i\left(\dfrac{x_i}{z-x_i}B_i^{(n)}(x)+\dfrac{q^nx_i}{z+q^nx_i}B_i^{(-n)}(x)\right),
\end{split}
\end{align}
at $z=0$ and $z=\infty$. We find that $S_0^B(x)=0$ and $S_1^B(x)=(1+q^n)p_1(x)$. Thus, we recover indeed the q-difference equation \ref{qdiff_p1} from equ. \ref{equ_p1}.

\subsection{C-case}
Once again, we exploit the fact that the operators $W_{0,n}^C$ are diagonal on the states $\ketC{\l}$ and coincide with the zero modes of the currents $w_n^C(z)$.
\begin{widetext}
From the expression $w_n^C(z)=-(1-q^n)z\hphi(z)\hphi(-q^nz)$, we compute the commutator
\begin{align}
\begin{split}
[w_n^C(z),\vdots\hphi(-x_1^{-1})\cdots\hphi(-x_N^{-1})\vdots]=\ &(1-q^n)q^{-n}\sum_{i=1}^N\d(-q^nx_iz)\prod_{j\neq i}(-x_i^{-1}-x_{j}^{-1})\ \hphi(z)\vdots\hphi(-x_1^{-1})\cdots\cancel{\hphi(-x_i^{-1})}\cdots\hphi(-x_N^{-1})\vdots\\
&-(1-q^n)\sum_{i=1}^N\d(x_iz)\prod_{j\neq i}(-x_i^{-1}-x_{j}^{-1})\ \vdots\hphi(-x_1^{-1})\cdots\cancel{\hphi(-x_i^{-1})}\cdots\hphi(-x_N^{-1})\vdots\hphi(-q^nz).
\end{split}
\end{align}
Using the first and second properties, we can insert $\hphi(z)$ and $\hphi(-q^nz)$ inside the symbols $\vdots\cdots\vdots$. Doing so, we reintroduce a dependence on the coordinate $x_i$ inside the correlator, but the argument of the corresponding field has an extra factor of $q^{\pm n}$. Thus, we end up with
\begin{align}
\begin{split}
&[w_n^C(z),\vdots\hphi(-x_1^{-1})\cdots\hphi(-x_N^{-1})\vdots]\\
=\ &(1-q^n)\sum_{i=1}^N\left(\d(-q^nx_iz)q^{-n}\prod_{j\neq i}\dfrac{x_i+x_j}{x_i+q^{-n}x_j}\ T_{q^n,x_i}-\d(x_iz)\prod_{j\neq i}\dfrac{x_i+x_j}{x_i+q^{n}x_j}T_{q^{-n},x_i}\right)\vdots\hphi(-x_1^{-1})\cdots\hphi(-x_N^{-1})\vdots.
\end{split}
\end{align}
Integrating over $z$, we deduce the commutator of the zero mode
\begin{align}
\begin{split}
[W_{0,n}^C,\vdots\hphi(-x_1^{-1})\cdots\hphi(-x_N^{-1})\vdots]=\ &\sum_{i=1}^N\left(q^{-n}\prod_{j\neq i}\dfrac{x_i+x_j}{x_i+q^{-n}x_j}\ T_{q^n,x_i}-\prod_{j\neq i}\dfrac{x_i+x_j}{x_i+q^{n}x_j}T_{q^{-n},x_i}\right)\vdots\hphi(-x_1^{-1})\cdots\hphi(-x_N^{-1})\vdots.
\end{split}
\end{align}
Projecting on the states $\braC{\vac}$ and $\ketC{\l}$, and using the vertical action \ref{WC_vert}, we deduce the q-difference equation
\begin{align}
\begin{split}
&\sum_{i=1}^N\left(\prod_{j\neq i}\dfrac{x_i+x_j}{x_i+q^{n}x_j}T_{q^{-n},x_i}-q^{-n}\prod_{j\neq i}\dfrac{x_i+x_j}{x_i+q^{-n}x_j}\ T_{q^n,x_i}\right)\braC{\vac}\vdots\hphi(-x_1^{-1})\cdots\hphi(-x_N^{-1})\vdots\ketC{\l}\\
=&\left((1-q^{n/2})q^{-n/2}\sum_{\sAbox\in\l}(\hchi_{\sAbox}^n+q^{-n/2}\hchi_{\sAbox}^{-n})\right)\braC{\vac}\vdots\hphi(-x_1^{-1})\cdots\hphi(-x_N^{-1})\vdots\ketC{\l}.
\end{split}
\end{align}
It only remains to multiply by the factor $\prod_ix_i^{K_N}$ to we find the q-difference equation \ref{C_qdiff}.
\end{widetext}

\bibliographystyle{apsrev4-2}
\bibliography{../BCD_Vertical}
\end{document}